\documentclass{article}
\usepackage{kr}

\usepackage{times}
\usepackage{soul}
\usepackage{url}
\usepackage[hidelinks]{hyperref}
\usepackage[utf8]{inputenc}
\usepackage[small]{caption}
\usepackage{graphicx}
\usepackage{amsmath}
\usepackage{amsthm}
\usepackage{booktabs}
\usepackage{algorithm}
\usepackage{xcolor}
\usepackage{tikz}
\usepackage[noend]{algpseudocode}
\usetikzlibrary{shapes.geometric}
\usepackage{xargs}

\usepackage{xspace}

\usepackage{xstring}

\usepackage[cmex10]{mathtools}
\usepackage{amssymb}
\usepackage{amsfonts}
\usepackage{mathrsfs}
\usepackage{latexsym}
\usepackage{textcomp}
\usepackage{pifont}

\newcommand{\argemp}[2]{\if&#1&\else#2\fi}

\newcommand{\argdef}[2]{\if&#1&#2\else#1\fi}

\newcommand{\argint}[3]{\if&#2&\else#1#2#3\fi}

\newcommand{\argext}[3]{\if&#1&#3\else#1\if&#3&\else#2#3\fi\fi}

\newcommandx{\mthfnt}[3][1=, 2=0]{{
	\IfStrEqCase{#1}
	{%
		{}%
		{#3}%
		{Name}%
		{%
			\IfStrEqCase{#2}
			{%
				{0}{\mathcal{#3}}%
				{1}{\mathscr{#3}}%
				{2}{\mathfrak{#3}}%
				{3}{\mathbb{#3}}%
			}
			[\ensuremath{\clubsuit}]%
		}%
		{Set}%
		{%
			\IfStrEqCase{#2}
			{%
				{0}{\mathrm{#3}}%
				{1}{\mathsf{#3}}%
				{2}{\mathbb{#3}}%
				{3}{\mathbf{#3}}%
			}
			[\ensuremath{\clubsuit}]%
		}%
		{Fun}%
		{%
			\IfStrEqCase{#2}
			{%
				{0}{\mathsf{#3}}%
				{1}{\mathrm{#3}}%
			}
			[\ensuremath{\clubsuit}]%
		}%
		{Rel}%
		{%
			\IfStrEqCase{#2}
			{%
				{0}{\mathit{#3}}%
				{1}{\mathtt{#3}}%
			}
			[\ensuremath{\clubsuit}]%
		}%
		{Sym}%
		{%
			\IfStrEqCase{#2}
			{%
				{0}{\mathtt{#3}}%
				{1}{\mathbf{#3}}%
			}
			[\ensuremath{\clubsuit}]%
		}%
		{Elm}%
		{\mathnormal{#3}}
	}
[\ensuremath{\clubsuit}]%
}}

\newcommand{\mthsub}[1]{\argemp{#1}{\ensuremath{_{\mathnormal{#1}}}}}

\newcommand{\mthsup}[1]{\argemp{#1}{\ensuremath{^{\mathnormal{#1}}}}}

\newcommandx{\mth}[5][1=, 2=0, 4=, 5=]{{\ensuremath{\mthfnt[#1][#2]{#3}\mthsub{#4}\mthsup{#5}}}}

\newcommandx{\mtharg}[6][1=, 2=0, 4=, 5=]{{\mth[#1][#2]{#3}[#4][#5]\ensuremath{\argint{(}{#6}{)}}}}

\newcommand{\mthstyname}{0}
\newcommand{\mthname}[1][]{\mth[Name][\argdef{#1}{\mthstyname}]}

\newcommand{\mthstyset}{0}
\newcommand{\mthset}[1][]{\mth[Set][\argdef{#1}{\mthstyset}]}

\newcommand{\mthstyfun}{0}
\newcommand{\mthfun}[1][]{\mth[Fun][\argdef{#1}{\mthstyfun}]}

\newcommand{\tuple}[1]
{\ensuremath{\!\argint{\langle}{#1}{\rangle}}}

\def\Nat{\mathbb{N}}

\newcommand{\pow}[1]{\ensuremath{2^{#1}}}

\newcommand{\LTL}{\mthfun{LTL}\xspace}

\def\TEMPORAL#1{\mbox{\small\boldmath$\mathbf{#1}$}}
\def\ltlnext{\TEMPORAL{X}}
\def\sometime{\TEMPORAL{F}} 
\def\always{\TEMPORAL{G}}
\def\until{\,\TEMPORAL{U}\,}
\def\Nat{\mathbb{N}} 

\newcommand{\Ag}{\mthset{N}}
\newcommand{\Ac}{\mthset{Ac}}
\newcommand{\AcProf}{\vec{\Ac}}
\newcommand{\St}{\mthset{St}}
\newcommand{\AP}{\mthset{AP}}
\newcommand{\APSet}{\AP}
\newcommand{\Pun}{\text{Pun}}

\newcommand{\Par}{\mthset{PAR}}

\newcommand{\CGModel}{\mthname{M}}
\renewcommand{\Game}{\mthname{G}}

\newcommand{\ParGame}{\Game_{\Par}}
\renewcommand{\Pun}{\mthset{Pun}}

\newcommand{\Automaton}{\mthname{A}}

\newcommand{\labFun}{\lambda}

\newcommand{\coreset}{\mathit{core}}

\newcommand{\WinSet}{\mthset{Win}}

\newcommand{\trnFun}{\mthfun{tr}}

\newcommand{\StrSet}{\Sigma}

\newcommand{\strElm}{\sigma}

\newcommand{\NE}{\mthset{NE}}

\newcommand{\winsym}{\mthset{Win}}
\newcommandx{\Win}[3][1=, 2=, 3=]
{\mthset{\winsym#3}[#1][#2]}

\newcommand{\presym}{\mthfun{Pre}}
\newcommandx{\Pre}[3][1=, 2=, 3=]
{\mthset{\presym#3}[#1][#2]}

\newcommand{\eqsym}{\mthfun{Eq}}
\newcommandx{\Eq}[3][1=, 2=, 3=]
{\mthset{\eqsym#3}[#1][#2]}

\newcommandx{\AFW}[5][1=, 2=, 3=, 4=, 5=]
{\txtargname{AFW#5{\small\argint{$[$}{#1}{$]$}}}[#2][#3]{#4}\xspace}

\def\exptime{\mthfun{EXPTIME}\xspace}
\def\twoexptime{\mthfun{2EXPTIME}\xspace}

\def\twoexptimeC{\mthfun{2EXPTIME}-complete\xspace}

\def\pspace{\mthfun{PSPACE}\xspace}

\def\np{\mthfun{NP}\xspace}

\newcommand{\MC}{\mathcal{C}}

\newcommand{\MDP}{\mathcal{K}}

\newcommand{\D}{\mthfun{D}}

\def\Rat{\mathbb{Q}}

\newcommand{\spt}{\mthfun{spt}}

\newcommand{\paths}{\mthset{Paths}}

\newcommand{\fpaths}{\mthset{Fpaths}}

\newcommand{\cyl}{\mthset{Cyl}}

\newcommand{\AS}{\mthfun{AS}}

\newcommand{\NZ}{\mthfun{NZ}}

\renewcommand{\phi}{\varphi}

\newcommand{\act}{a}
\newcommand{\jact}{\vec{a}}

\algnewcommand\Input{\item[ ]{\textbf{input:} }}
\algrenewcommand\Return{\item{\textbf{return} }}

\newcommand{\proj}{\mthfun{proj}}

\newtheorem{example}{Example}
\newtheorem{theorem}{Theorem}
\newtheorem{lemma}{Lemma}
\newtheorem{proposition}{Proposition}
\theoremstyle{definition}
\newtheorem{definition}{Definition}
\theoremstyle{remark}

\title{Rational Verification for Probabilistic Systems}

\author{%
Julian Gutierrez$^1$\!\!\and\!\!\!
Lewis Hammond$^2$\!\!\and\!\!\!
Anthony W. Lin$^3$\!\!\and\!\!\!
Muhammad Najib$^3$\!\!\and\!\!\!
Michael Wooldridge$^2$ \\
\affiliations
$^1$Monash University\\
$^2$University of Oxford\\
$^3$University of Kaiserslautern\\
\emails
julian.gutierrez@monash.edu, 
\{lewis.hammond, mjw\}@cs.ox.ac.uk, 
\{lin, najib\}@cs.uni-kl.de
}

\begin{document}

\maketitle

\begin{abstract}
  % Length: 9 pages, including abstract, figures, and appendices
  % (if any), but excluding references and acknowledgements. The abstract should be no more than 200 words long. Authors may optionally provide supplementary material (e.g. proof details, additional experimental results)
  % as a separate file. Such material will be consulted at the discretion of reviewers and will not be published.
  % Please refer to the KR2021website\footnote{\url{https://kr2021.kbsg.rwth-aachen.de/page/call_for_papers}}
  % for further information. Illustrations/tables/algorithms should be floated to the top (preferred) or bottom of the page, unless
  % they are an integral part of your narrative flow. When placed at the bottom or
  % top of a page, illustrations may run across both columns, but not when they
  % appear inline. Captions should always appear below the illustration.
  Rational verification is the problem of determining which temporal logic
    properties will hold in a multi-agent system, under the assumption that
    agents in the system  act rationally, by choosing strategies that
    collectively form a game-theoretic equilibrium. Previous work in this area
    has largely focussed on deterministic systems. In this paper, we develop the
    theory and algorithms for rational verification in probabilistic systems. We
    focus on \emph{concurrent stochastic games} (CSGs), which can be used to
    model uncertainty and randomness in complex multi-agent environments. We study the
    rational verification problem for both non-cooperative games and cooperative
    games in the qualitative probabilistic setting. In the former case, 
    we consider \LTL properties satisfied by the Nash equilibria of the game and in the latter case \LTL properties satisfied by the core. In both cases, we show that the problem is \twoexptimeC, thus not harder than the much simpler verification problem of model checking \LTL properties of systems modelled as Markov decision processes (MDPs).
\end{abstract}

\section{Introduction}

\emph{Rational verification} is the problem of determining which temporal logic properties will hold in a multi-agent system, under the assumption that agents in the system act rationally, by choosing strategies/policies for acting which collectively form a game-theoretic equilibrium~\cite{GutierrezHW17,WooldridgeGHMPT16}. Rational verification has been studied for a range of models: typically, each agent is modelled as a non-deterministic reactive program, where non-determinism captures the choices available to each agent at each time step -- the strategies available to a player correspond to each possible way that an agent can resolve its non-determinism. To be able to reason about game-theoretic equilibria, the model also needs to capture the preferences that players have, and a common approach for this is to associate with each agent a temporal logic ``goal'' formula that the player desires to be satisfied. For example, in the Reactive Modules reasoning framework, agents are modelled using the Reactive Modules language~\cite{alur:99a}, and agent preferences are modelled with goals expressed in linear temporal logic (\LTL)~\cite{GutierrezNPW20,GutierrezNPW18}. 

The most basic decision problem in rational verification is as follows: \emph{Given a system $M$, and a temporal logic formula $\phi$, does there exist a Nash equilibrium profile of strategies $\vec{\sigma} = (\sigma_1, \ldots, \sigma_n)$ for the players in $M$ such that $\phi$ will be satisfied under the assumption that players act according to $\vec{\sigma}$}. This decision problem is known as \textsc{E-Nash}, and is \twoexptime-complete for Reactive Modules games~\cite{GutierrezHW17}; the corresponding \textsc{A-Nash} problem asks, instead, whether formula~$\phi$ holds for \emph{all} Nash equilibrium profiles in the game. 

Although many models have been studied in the context of rational verification, little research has considered \emph{probabilistic} models, and as such, existing models are limited in the scope of domains they can capture. Our aim in this paper is to rectify this omission: we study rational verification in probabilistic systems in which players have goals represented by \LTL formulae. Our basic model is called \emph{concurrent stochastic games} (CSGs), sometimes also referred to as \emph{Markov games}. As in conventional concurrent games, a game is played over an infinite sequence of rounds, and at each round, every player chooses an action to perform. Unlike conventional concurrent games, however, the performance of a profile of actions does not induce a unique successor state, but rather a probability distribution over possible successor states. The main difference between the games we study and standard CSGs is that player preferences in our setting are defined by associating \LTL goals $\gamma_i$ with each player $i$. The game is played by each player choosing a strategy (cf.~\emph{policy}), which defines how that player will make choices over time when playing the game.  

In common with previous work, we model strategies as state machines with output (although we may require them to have infinite memory), though strategies in our setting are not required to choose a unique action at every time step, but instead choose a probability distribution over possible actions. A CSG together with a profile of strategies induces a Markov chain (MC), and given such an MC, we can determine the probability of given temporal formulae being satisfied, and in particular, the probability with which goal formulae $\gamma_i$ are satisfied. 
In this paper, we consider rational verification of CSGs in the \emph{qualitative} setting, {\em i.e.}, where we are interested in checking if the probability of satisfying \LTL goals is $1$ or greater than $0$. These are also known as 
%To deal with this situation, we adopt an idea from the verification of stochastic systems: we focus on the 
\emph{almost-sure} ($\AS$) and \emph{non-zero} ($\NZ$) satisfaction respectively, which -- together with their negated formulations, probability less than $1$ or equal to $0$ -- form one of the
most useful settings in probabilistic  verification \cite{BK08}. For example, they
can be used to
capture interesting liveness properties for many randomised protocols 
% in which there is some probability of a message being lost or corrupted, and where our concern is in ensuring that eventually a message is received by some node(s) in the network
\cite{Norman04,LR81,fairy-tale,LR16}.

Against this background, we study the rational verification problem for both cooperative and non-cooperative solution concepts. In the {\em non-cooperative} case, we focus on {\em Nash equilibria}, and characterise the complexity of the \textsc{E-Nash} and \textsc{A-Nash} problems, together with related problems. We then investigate \emph{cooperative} solution concepts, adapting the model of the \emph{core} that was introduced by \cite{GKW2019}, which defines strategy profiles that are stable against beneficial deviations by \emph{groups} of players (\emph{coalitions}), as opposed to individual deviations in the case of Nash equilibria.
% which can be viewed as the set of strategy profiles from which no \emph{coalition} (as opposed to \emph{player} in the case of Nash equilibria) has a beneficial deviation. 
We consider \textsc{E-Core},  \textsc{A-Core}, and related decision problems -- which in previous work have been studied only for concurrent multi-agent systems without any probabilistic behaviour.

To the best of our knowledge, this is the first work that considers the rational verification problem for probabilistic systems in which {\em games} can be cooperative or non-cooperative, players' {\em preferences} can be expressed using general \LTL goals, strategies may have access to {\em infinite} memory, interactions can take place {\em concurrently}, and plays may last for an infinite number of rounds ({\em i.e.}, have an {\em infinite horizon}). All previous work, in the probabilistic setting, fails to have at least one of these features arising in full interplay with the others, making our framework the most complex so far developed from a theoretical point of view. 
Indeed, because of the many features we consider together, several new constructions and proof techniques are required to be able to fully account for probabilistic behaviour.

\subsubsection{Structure of the paper}
In Section~\ref{secn:prelim}, the necessary background is given. Sections~\ref{secn:noncoop} and~\ref{secn:coop} contain the main technical results for non-cooperative games and cooperative games respectively. Then, in Section~\ref{secn:conc}, we present some concluding remarks and a summary of relevant related work.
 
% \begin{table}[h]
% 	\centering
% 	\begin{tabular}{l c c}
% 		\toprule
% 		\textbf{Problem} & \textbf{Lower Bound} & \textbf{Upper Bound} \\
% 		\midrule
% 		\textsc{Membership} & \twoexptime & \twoexptime \\
% 		\textsc{Non-Emptiness} &  &  \\
% 		\textsc{E-Nash} & \twoexptime & \twoexptime \\
% 		\textsc{A-Nash} & \twoexptime & \twoexptime \\
% 		\midrule
% 		\textsc{Core Membership} &  &  \\
% 		\textsc{Beneficial Deviation} &  &  \\
% 		\textsc{E-Core} &  &  \\
% 		\textsc{A-Core} &  &  \\
% 		\bottomrule
% 	\end{tabular}
% 	\caption{Summary of complexity results so far}
% \end{table}

%\newpage 
\section{Preliminaries}
\label{secn:prelim}
%\begin{itemize}
%\item CSGs
% \item Multiplayer games  
%\item Logics 
%\item Probabilistic model checking 
%\item Nash equilibria and the core
%\end{itemize}

For a finite set $ X $, a (rational) \textit{probability distribution} over $ X $ is a function $ \Pr : X \to [0,1] \cap \Rat $ such that $ \Sigma_{x \in X} \Pr(x) = 1 $. We write $ \D(X) $ for the set of probability distributions on $ X $, and $ \spt(\Pr) = \{ x \in X : \Pr(x) > 0 \} $ for the \textit{support} of the distribution $ \Pr $ on $ X $. For a tuple $ \vec{x} = (x_1,\dots,x_n) $, we write $ \proj_{i}(\vec{x}) = x_i $, {\em i.e.}, its $ i $-th \textit{projection}, and also $ \proj_{x_i}(\vec{x}) = x_i $  when the context is clear.

\subsubsection{Markov chains} 
A (discrete time) \textit{Markov chain} (MC) is a tuple $ \MC = (S, s_{\iota}, \trnFun,\lambda) $, where $ S $ is a set of states, $ s_{\iota} $ is the initial state, 
%$ E \subseteq \St \times \St $ is a set of edges, 
$ \trnFun : S \to \D(S) $ is a function that assigns a probability distribution (on the set of states $ S $) to all states $ s \in S $, and $ \lambda : S \to 2^{\AP} $ is a labelling function mapping each state to a set of propositions taken from the set $ \AP $.

The set of infinite paths in $ \MC $ starting from $ s \in S $ is $ \paths(\MC, s) = \{ \pi = s_0s_1\dots \in \St^{\omega} : s_0 = s, \forall k \in \Nat.~ \trnFun(s_k,s_{k+1}) > 0 \} $. The set of all infinite paths in $ \MC $ is $ \paths(\MC) = \bigcup_{s \in S} \paths(\MC,s) $. 
% By $ \pi_k $ we refer to the $ (k+1) $-th state in $ \pi $, and by $ \pi_{\geq k} $ to the (finite) prefix of $ \pi $ up to the $ (k+1) $-th element. 
The set of finite paths starting from $ s \in S $ is defined as $ \fpaths(\MC,s) = \{ \hat{\pi} = s_0 \cdots s_n \in S^+ : \exists \pi \in \paths(\MC).~ \hat{\pi}\pi \in \paths(\MC,s) \} $ and $ \fpaths(\MC) = \bigcup_{s \in S} \fpaths(\MC,s) $. 
The \textit{cylinder set} of a finite path $ \hat{\pi} \in \fpaths(\MC) $ is defined by $ \cyl(\hat{\pi}) = \{ \pi \in \paths(\MC) : \exists \tilde{\pi} \in \paths(\MC).~ \pi = \hat{\pi}\tilde{\pi} \in \paths(\MC) \} $. Following \cite{vardi85}, we define the probability distribution over the space of infinite paths, as usual, via cylinder sets. We denote this probability distribution over the set of infinite paths beginning from some state $ s $ by $ \Pr^s_{\MC} $. We also write $ \Pr_{\MC} $ when $ s $ is clear from the context.

% \subsubsection{Markov decision processes}
% A finite \textit{Markov decision process} (MDP) is a tuple $ \MDP = (S,s_{\iota},\Ac,\trnFun,\lambda) $, where $ S $ is a finite set of states, $ s_{\iota} $ is the initial state, $ \Ac $ is a finite set of actions, 
% %$ E \subseteq S \times \Ac \times S $ is a set of edges,
% $ \trnFun: S \times \Ac \to \D(S) $ is a function that assigns probability distribution (on the set of states $ S $) to all states $ s \in S $ when action $ a \in \Ac $ is chosen in $ s $, and $ \lambda : S \to 2^{\AP} $ is a labelling function. (Note that we do not use reward functions, since preferences will be defined by \LTL goal formulae.)

\subsubsection{Concurrent stochastic game arenas} 
A \textit{concurrent stochastic game arena} (CSGA) is a tuple $ \CGModel = (\Ag, \St, s^0, (\Ac_i)_{i \in \Ag}, \trnFun) $, where $ \Ag $ is a finite set of players, $ \St $ is a finite set of states, $ s^0 $ is the initial state, $ \Ac_i $ is a finite set of actions for each $ i \in \Ag $. With each player~$ i $ and each state $ s \in \St $, we associate a non-empty set $ \Ac_i(s) $ of \textit{available} actions that, intuitively, $ i $ can perform when in state $ s $. When all players have fixed their actions, we have an action profile $ \vec{a} = (a_1,\dots,a_n) \in \AcProf = \Ac_1 \times \cdots \times \Ac_n $ which we refer as as a \textit{direction}. A direction $ \vec{a} $ is available in state $ s $ if for all $ i $ we have $ a_{i} \in \Ac_{i}(s) $. We write $ \AcProf(s) $ for the set of available directions in state $ s $. 

For a given set of players
$A \subseteq \Ag$ and an action profile $\vec{a}$, we let
$\vec{a}_{A}$ and $\vec{a}_{-A}$ be two tuples of actions,
respectively, one for each player in $A$ and one for each player in
$\Ag \setminus A$. 
% We also write $\vec{a}_{i}$ for $\vec{a}_{\{i\}} $
% and $ \vec{a}_{-i} $ for $ \vec{a}_{\Ag \setminus \{i\}} $.
Furthermore,
for two directions $\vec{a}$ and $\vec{a}'$, we write
$(\vec{a}_{A}, \vec{a}_{-A}')$ to denote the direction where the
actions for players in $ A $ are taken from $\vec{a}$ and the actions
for players in $ \Ag \setminus A $ are taken from $\vec{a}'$. Finally, $ \trnFun: \St \times \AcProf \to \D(\St) $ is a probabilistic transition function. A Markov decision process (MDP), without a reward function, is simply a CSGA with one player only.

\subsubsection{Linear temporal logic}
\LTL~\cite{pnueli:77a} extends classical propositional logic with two
operators, $\ltlnext$ (``next'') and $\until$ (``until''), which can be
used to express properties of paths.  The syntax of \LTL is defined
with respect to a set $\APSet$ of propositional variables by the following grammar:
$$ \phi ::=
\mathop\top \mid
p \mid
\neg \phi \mid
\phi \vee \phi \mid
\ltlnext \phi \mid
\phi \until \phi 
$$
where $p \in \APSet$. Other connectives are defined in terms of $ \lnot $ and $ \vee $ in the usual way. Two key derived \LTL operators are $ \sometime $ (``eventually'') and $ \always $ (``always''), which are defined in terms of $ \until $ as follows: $ \sometime \phi \equiv \top \until \phi $ and $ \always \phi \equiv \lnot \sometime \lnot \phi $.

We interpret formulae of \LTL with respect to triples $(\pi,t,\labFun)$, where
$\pi \in \St^\omega$ is a path, $t \in \Nat$ is a
temporal index into $\pi$, and $\labFun: \St \to \pow{\APSet}$ is a
labelling function that indicates which propositional variables are
true in every state.  The semantics of \LTL is
given by the following rules:

\vspace*{1.5ex}
\noindent $\small \hspace*{-1.5ex}\begin{array}{lcl}
(\pi,t,\labFun)\models\mathop\top	\\
(\pi,t,\labFun)\models p 				&\text{ iff }&	p\in\labFun(\pi_t)\\
(\pi,t,\labFun)\models\neg \phi			&\text{ iff }&   \text{it is not the case that $(\pi,t,\labFun) \models \phi$}\\
(\pi,t,\labFun)\models\phi \vee \psi		&\text{ iff }&	\text{$(\pi,t,\labFun) \models \phi$  or $(\pi,t,\labFun) \models \psi$}\\
(\pi,t,\labFun)\models\ltlnext\phi			&\text{ iff }&	\text{$(\pi,t+1,\labFun) \models \phi$}\\
(\pi,t,\labFun)\models\phi\until\psi	&\text{ iff }&   \text{for some $t' \geq t: \ \big((\pi,t',\labFun) \models \psi$  and }\\
&&\text{for all $t \leq t'' < t': \ (\pi,t'',\labFun) \models \phi \big)$}\\
\end{array}
$
\vspace*{0.5ex}

\noindent If $(\pi,0,\labFun)\models\phi$, we write $\pi\models\phi$ and say that
\emph{$\pi$ satisfies~$\phi$}. 

\subsubsection{Concurrent stochastic games}
A \textit{concurrent  stochastic game} (CSG) is a tuple $ \Game = (\CGModel, (\gamma_i)_{i \in \Ag}, \lambda) $, where $ \CGModel $ is a CSGA, $ \gamma_i $ is a \LTL formula that represents the \textit{goal} of player~$ i $, and $ \lambda : \St \to 2^{\AP} $ a labelling function. 
A game is played by each player~$i$ selecting a \emph{strategy}~$\sigma_i$
that defines how it makes choices over time. 
A strategy for player $ i $ can be understood as a function $ \sigma_i : \St^{+} \to \D(\Ac_i) $ that assigns to every non-empty finite sequence of states a probability distribution over player~$i$'s set of actions. In general, strategies require memory to remember the history of the game. When a strategy remembers a finite amount of information about the past we call it \textit{finite-memory}, and when each distribution $\sigma_i(s^{+})$ is deterministic we call $\sigma_i$ \textit{pure}.

Formally, a \textit{strategy} in $ \Game $ for player~$ i $ is a a transducer $\strElm_{i} = (Q_{i}, q_{i}^{0}, \delta_i, \tau_i) $, where $Q_{i}$ is a (possibly infinite) set of \emph{internal states}, $ q_{i}^{0} $ is
the \emph{initial state},
$\delta_i: Q_{i} \times \St \rightarrow Q_{i} $ is a deterministic
\emph{internal transition function}, and
$\tau_i: Q_{i} \times \St \rightarrow \D(\Ac_i) $ an \emph{action function} that selects a distribution on $ \Ac_i $ such that for all $ q_i \in Q_i $ and $ s \in \St $, we have $ \tau_i(q_i,s) \in \D(\Ac_i(s)) $. Let $\StrSet_i$ be the set of strategies for player $i$. A strategy is \textit{memoryless} if there exists a transducer encoding the strategy with $ |Q_i| = 1 $, {\em i.e.}, the choice of action only depends on the current state of the game, and finite-memory if $ |Q_i| < \infty $. Moreover, a strategy is said to be \textit{deterministic} if $ \tau_i : Q_i \times \St \to \Ac_i $, such that for every $ q_i \in Q_i $ and every $ s \in \St $, we have that $ \tau_i(q_i,s) \in \Ac_i(s) $.

Once every player~$i$ has selected a strategy~$\sigma_i$, we have a 
\emph{strategy profile}~$\vec{\sigma} = (\sigma_1, \dots, \sigma_n)$. We write $ \vec{\sigma}_{A} $ and $ \vec{\sigma}_{-A} $ to denote the strategy profile for players in $ A \subseteq \Ag $ and $ N \setminus A $, respectively. We also write $ (\vec{\sigma}_{A}, \vec{\sigma}_{-A}') $ to denote the strategy profile where the strategies for players in $ A $ are taken from $ \vec{\sigma} $, and the strategies for players in $ \Ag \setminus A $ are taken from $ \vec{\sigma}' $.
Observe that a strategy profile $ \vec{\sigma} $ for a game $ \Game $ resolves nondeterminism in the underlying $ \CGModel $. That is, a strategy profile $ \vec{\sigma} $ for a game $ \Game $ induces an MC $ \MC_{\vec{\sigma}} = (S,s_\iota,\trnFun',\lambda) $, 
%\mnnote{still working on this definition\dots}
where $ S = \St \times \bigtimes_{i \in \Ag} Q_i $, $ s_\iota = (s^0,q^0_1,\dots,q^0_n) $, and for $ v,v' \in S, \trnFun'(v,v') = \sum_{\vec{a} \in \AcProf} \prod_{\vec{a}_i \in \vec{a}} \tau_i(\proj_{q_i}(v),\proj_s(v),\vec{a}_{i}) \cdot \trnFun(\proj_s(v),\vec{a},\proj_s(v')) $, if for each $ i \in \Ag $, $ \proj_{q_i}(v') = \delta_i(\proj_{q_i}(v),\proj_s(v)) $, and is not defined otherwise.

%\newpage 
\subsubsection{Automata}
A \emph{deterministic automaton on infinite words} is given by a structure
$\Automaton = (\AP, Q, q^0, \rho, \mathcal{F}) $, where $ Q $ is a
finite set of states, $ \rho: Q \times \AP \rightarrow Q $ is a
transition function, $ q^0 $ is an initial state, and $\mathcal{F}$ is an
acceptance condition. A \textit{parity} condition $ \mathcal{F} $ is a partition
$ \{F_1,\dots,F_n\} $ of $ Q $, where $ n $ is the \emph{index} of the
parity condition and any $ k \in [1,n] $ is a \emph{priority}. We use
a \emph{priority function} $ \alpha: Q \rightarrow \mathbb{N} $ that maps
states to priorities such that $ \alpha(q)=k $ if and only if
$ q \in F_k $. For a run $ \pi = q^0q^1q^2\dots $, let
$ \mathit{inf}(\pi)$ denote the set of states occurring infinitely often
in the run $ \mathit{inf}(\pi) = \{q \in Q : q = q^i \text{ for infinitely many \emph{i}'s}\} $. A run $\pi$ is accepted by a deterministic
parity word (DPW) automaton with condition $ \mathcal{F} $ if the
minimum priority that occurs infinitely often is even, {\em i.e.}, if $\left(\min_{k \in [1,n]}(\mathit{inf}(\pi) \cap F_k \neq \varnothing)\right)
\bmod 2 = 0.$
%A Streett condition $ \mathcal{F} $ is a set of pairs
%$ \{(E_1,C_1),\dots,(E_n,C_n)\} $ where $ E_k \subseteq Q$ and
%$ C_k \subseteq Q $ for all $ k \in [1,n] $. A run $ \pi $ is
%accepted by a deterministic Streett word (DSW) automaton~$\mthname{S}$
%with condition $ \mathcal{F} $ if  $\pi$
%either visits $ E_k $ finitely many times or visits $ C_k $ infinitely
%often, i.e., if for every $ k $ either
%$ \mathit{inf}(\pi) \cap E_k = \varnothing $ or
%$ \mathit{inf}(\pi) \cap C_k \neq \varnothing $.

For a given game $ \Game $ and a strategy profile $ \vec{\sigma} $, a formula $ \varphi $ is said to be \textit{almost-surely} satisfied, denoted $ \vec{\sigma} \models \AS(\varphi) $, if and only if, $ \Pr_{\MC_{\vec{\sigma}}}(\{ \pi \in \paths(\MC_{\vec{\sigma}},s^0) : \pi \models \phi \}) = 1 $. Similarly, we say that $ \varphi $ is satisfied with \emph{non-zero} probability, denoted $ \vec{\sigma} \models \NZ(\varphi) $ if $ \Pr_{\MC_{\vec{\sigma}}}(\{ \pi \in \paths(\MC_{\vec{\sigma}},s^0) : \pi \models \phi \}) > 0 $. Observe that $\NZ$ can be viewed as the dual of $\AS$, written (with a slight abuse of notation) as $ \NZ(\varphi) \equiv \neg \AS (\neg \varphi)$. Hence, for ease of exposition, in the remainder of the paper we focus on $\AS$ winning conditions, with the understanding that all our results from this case can be also used in the case of $\NZ$ winning conditions, and their respective negated formulations.

A \textit{concurrent multiplayer stochastic parity game} (CSPG) is given by a structure 
	$ \ParGame = (\CGModel,(\alpha_i)_{i \in \Ag}) $ 
	where $ \alpha_i : \St \to \Nat $ is the goal of player $ i $, given as a priority function over the set of states $ \St $. A path $ \pi $ satisfies a priority function $ \alpha $, denoted by $ \pi \models \alpha $, if the minimum number occuring infinitely often in the infinite sequence $ \alpha(\pi_0)\alpha(\pi_1)\alpha(\pi_2)\dots $ is even. 
	Almost-surely satisfaction in CSPGs is then defined in a similar way: we say that 
	$ \vec{\sigma} \models \AS(\alpha) $  if and only if $ \Pr_{\MC_{\vec{\sigma}}}(\{ \pi \in \paths(\MC_{\vec{\sigma}},s^0) : \pi \models \alpha \}) = 1 $.

For a CSG $ \Game $, strategy profile $ \vec{\sigma} $, and state $ s $, we define the set of winners and losers by $ W_{\Game}(\vec{\sigma},s) = \{ i \in \Ag : (\vec{\sigma},s) \models \AS(\gamma_i) \} $ and $ L_{\Game}(\vec{\sigma},s) = \{ i \in \Ag : (\vec{\sigma},s) \models \neg \AS(\gamma_i) \} $. We also write $ W_{\Game}(\vec{\sigma}) $ and $ L_{\Game}(\vec{\sigma}) $, for $ W_{\Game}(\vec{\sigma},s^0) $ and $ L_{\Game}(\vec{\sigma},s^0) $. We define the above concepts for CSPGs analogously, with $\ParGame$ replacing $\Game$ and $\alpha_i$ replacing $\gamma_i$.

% \subsubsection{Preferences}
% Since the strategy profile played in a game determines if a player's goal is almost-surely satisfied, we can define a 
% preference relation~$\succeq_i$ over such profiles for each player~$i$. 
% %Let~$w_i$ be $\gamma_i$ if $\Game$ is an \LTL game, and be $\alpha_i$ if $\Game$ is a Parity game. 		
% %	Let~$w_i=\gamma_i$ and $w_i=\alpha_i$ if $\Game$ is an \LTL game or if $\Game$ is a Parity game, respectively. 
% Given $\vec\sigma$ and $\vec\sigma'$ in $\Game$, we have 
% $$
% \text{$\vec{\sigma} \succeq_i \vec{\sigma}'$\ ~~~iff~~~ \ 
% 	$\vec\sigma'\models \AS(\gamma_i) $ implies $\vec\sigma \models \AS(\gamma_i)$.}
% $$

%\newpage \dots 

%\newpage 
\section{Non-Cooperative Rational Verification}
\label{secn:noncoop}

We now introduce rational verification problems involving non-cooperative solution concepts -- and in particular, problems relating to \emph{Nash equilibria}~\cite{OR94}. We begin by defining this concept for our setting: Given a game $\Game$, a strategy profile $\vec{\sigma}$ is a \emph{Nash equilibrium} of~$\Game$ if, for every player~$i$ and strategy $\sigma'_i\in\Sigma_i$, we have 
$$(\vec{\sigma}_{-i},\sigma'_i)\models \AS(\gamma_i) ~\text{ implies }~ \vec\sigma \models \AS(\gamma_i)$$
\noindent where $(\vec{\sigma}_{-i},\sigma'_i)$ denotes $(\sigma_1, \dots, \sigma_{i - 1}, \sigma'_i, \sigma_{i + 1}, \dots, \sigma_n)$, the strategy profile where the strategy of player~$i$ in $\vec{\sigma}$ is replaced by $\sigma'_i$.
Note that this is equivalent to a more traditional formulation in which the utility function of each player $i$ is defined as equal to a constant $a$ if $\AS(\gamma_i)$ holds and equal to a constant $b < a$ otherwise.
Let $\NE(\Game)$ denote the set of Nash equilibria of~$\Game$.
We begin by introducing the key rational verification problems for non-cooperative settings; these are the natural adaptation of rational verification for the almost-sure setting.
% \vspace{4pt}

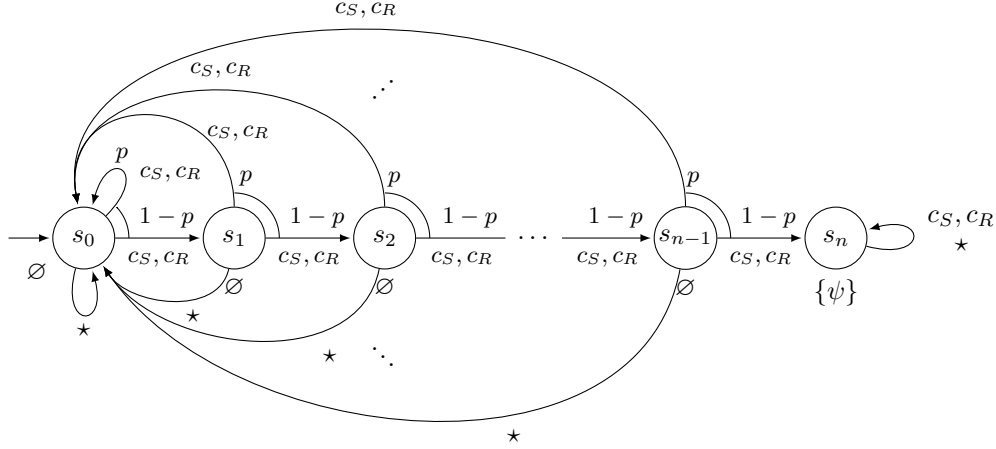
\begin{figure*}[hbt!]
	
	%	\begin{adjustbox}{width=\textwidth}
	\centering
	\begin{tikzpicture}
	[->,>=latex,shorten >=1pt,auto,node distance=1cm, auto, main node/.style={circle,draw, text width=8mm, align=center, inner sep=0pt}, sq node/.style={minimum size=0.5cm,draw}, no arrow/.style={-}]
	
	\node[main node, label={$  $}, label={[label distance=0.1]210:$\varnothing$}] (0) [] {$ s_0 $};
	\draw [domain=0:45,-] plot ({0.6*cos(\x)}, {0.6*sin(\x)});
	
	%		\node[sq node,  label={$ $}, label=below:{$ $}] (1) [right of=0, yshift=1cm,xshift=1cm] {};
	
	\node[main node, label={$  $}, label=below :{$ \varnothing $}] (1) [right of=0,xshift=1cm,yshift=0cm] {$ s_1 $};
	\draw [xshift=2cm, domain=0:93,-] plot ({0.6*cos(\x)}, {0.6*sin(\x)});
	
	\node[main node, label={$  $}, label=below:{$ \varnothing $}] (2) [right of=1,xshift=1cm,yshift=0cm] {$ s_2 $};
	\draw [xshift=4cm, domain=0:93,-] plot ({0.6*cos(\x)}, {0.6*sin(\x)});
	
	\node (3) at (6,0) {$ \dots $};
	
	\node[main node, label={$  $}, label=below:{$ \varnothing $}] (4) [right of=3,xshift=1cm,yshift=0cm] {$s_{n-1}$};
	\draw [xshift=8cm, domain=0:93,-] plot ({0.6*cos(\x)}, {0.6*sin(\x)});
	
	\node[main node, label={$  $}, label=below:{$ \{ \psi \} $}] (5) [right of=4,xshift=1cm,yshift=0cm] {$s_{n}$};
	
	\node[rotate=45] (6) at (4,2) {$ \dots $};
	\node[rotate=-45] (7) at (4,-1.6) {$ \dots $};
	
	%		\node[sq node,  label={}, label=below:{}] (2) [below of=0, yshift=-1cm] {};
	
	%		\node[main node, label={$  $}, label=below:{$\varnothing$}] (4) [right of=0,xshift=1cm,yshift=-1cm] {$ s_2 $};
	
	\path
	++ (-1,0) edge node[]{} (0)
	(0) edge [bend right=-0] node[xshift=0cm,yshift=-0.5cm]{\small $ c_S,c_R $} node[xshift=0.1cm,yshift=0.02cm]{\small $ 1-p $} (1)
	(1) edge [bend right=-0] node[xshift=0cm,yshift=-0.5cm]{\small $ c_S,c_R $} node[xshift=0.1cm,yshift=0.02cm]{\small $ 1-p $} (2)
	
	(2) edge [no arrow] node[yshift=-0.5cm]{\small $ c_S,c_R $} node[xshift=0.1cm,yshift=0.02cm]{ \small $ 1-p $} (3)
	
	(3) edge node[yshift=-0.5cm]{\small $ c_S,c_R $} node[xshift=0.1cm,yshift=0.02cm]{ \small $ 1-p $} (4)
	
	(4) edge [bend right=-0] node[xshift=0cm,yshift=-0.5cm]{\small $ c_S,c_R $} node[xshift=0.1cm,yshift=0.02cm]{\small $ 1-p $} (5)
	%		(1) edge [bend right] node[xshift=-0.2cm,yshift=0.55cm]{$ \frac{1}{2} $} (0)
	%		(0) edge [bend right=15] node[align=center,xshift=-0.7cm,yshift=-1.3cm]{$ a,\bar{b} $\\$ \bar{a},b $\\$ \bar{a},\bar{b} $} (4)
	
	(0) edge [loop above, looseness=10, in=75, out=45] node[]{\small $ p $} node[xshift=0.65cm,yshift=-0.25cm]{\small $ c_S,c_R $} (0)
	
	(0) edge [loop below, looseness=10, in=285, out=255] node[]{$ \star $ } node[xshift=0.65cm,yshift=-0.25cm]{} (0)
	
	(5) edge [loop above, looseness=10, in=15, out=-15] node[]{} node[align=center, xshift=0.65cm,yshift=-0.3cm]{$ c_S,c_R $\\$\star$} (5)
	%		(4) edge [loop right](4)
	%		(3) edge [loop right](3)
	
	(1) edge [bend right=0, looseness=2, in=280, out=270] node[xshift=1.2cm,yshift=0cm]{\small $ c_S,c_R $} node[xshift=1.3cm,yshift=-0.6cm]{\small $ p $} (0)
	(2) edge [bend right=0, looseness=1.3, in=280, out=270] node[xshift=0cm,yshift=0.5cm]{\small $ c_S,c_R $} node[xshift=2.3cm,yshift=-0.95cm]{\small $ p $} (0)
	(4) edge [bend right=0, looseness=1, in=280, out=270] node[xshift=0cm,yshift=0.5cm]{\small $ c_S,c_R $} node[xshift=4.35cm,yshift=-1.75cm]{\small $ p $} (0)
	
	(1) edge [bend right=0, looseness=1, in=125, out=80] node[xshift=0.5cm,yshift=0cm]{ $ \star $} node[xshift=1.3cm,yshift=-0.6cm]{\small} (0)
	(2) edge [bend right=0, looseness=1, in=125, out=80] node[xshift=1.2cm,yshift=0cm]{$ \star $} node[xshift=1.3cm,yshift=-0.6cm]{\small} (0)	
	(4) edge [bend right=0, looseness=1, in=125, out=80] node[xshift=1.2cm,yshift=0cm]{$ \star $} node[xshift=1.3cm,yshift=-0.6cm]{\small} (0)
	;
	
	\end{tikzpicture}
	%	\end{adjustbox}
	\caption{The CSGA for Examples~\ref{ex:1} and \ref{ex:2}. Edges without probability labels mean they have probability 1, and edges labelled $ \star $ are executed if the action profile is not equal to $ c_S,c_R $.}
	\label{fig:ex1ts}
\end{figure*}

\begin{quote}
	\underline{\textsc{Membership}} \\
	\emph{Given}: Game $\Game$, strategy profile $\vec{\sigma}$.\\
	\emph{Question}: Is it the case that $\vec{\sigma} \in \NE(\Game)$?
\end{quote}

\begin{quote}
	\underline{\textsc{E-Nash}} \\
	\emph{Given}: Game $\Game$, \LTL formula $\varphi$.\\
	\emph{Question}:  Is it the case that $\exists \vec{\sigma} \in \NE(\Game).\ \vec{\sigma} \models \AS(\varphi)$?
\end{quote}
\noindent
We can also ask the obvious counterpart of \textsc{E-Nash}:

\begin{quote}
	\underline{\textsc{A-Nash}} \\
	\emph{Given}: Game $\Game$, \LTL formula $\varphi$.\\
	\emph{Question}: Is it the case that $\forall \vec{\sigma} \in \NE(\Game).\ \vec{\sigma} \models \AS(\varphi)$?
\end{quote}

The intuitively simpler question of asking whether a game $\Game$ has any Nash equilibria, typically known as \textsc{Non-Emptiness} in the rational verification literature, can be solved simply by checking if $(\Game,\top) \in \mbox{\textsc{E-Nash}}$. Note that the question of \textsc{Non-Emptiness} may be non-trivial, as the fact that in our setting strategies can have infinite memory (and thus there are infinitely many of them) means we cannot straightforwardly apply Nash's theorem.

To illustrate some of the concepts introduced above, we will make use of the following example. 

\begin{example}\label{ex:1}
%	Consider a game with two players $ \Ag = \{1,2\} $ and two variables $ \AP =\{p,q\} $, with player~$ 1 $'s action set being $ \Ac_1 = \{a,\bar{a}\} $ and player~$ 2 $'s being $ \Ac_2 =\{b,\bar{b} \} $. The game is played on a CSGA shown in Figure~\ref{fig:ex1ts}, and moreover, let $ \gamma_1 = \sometime p $ and $ \gamma_2= \sometime q $. The objective for player~$ i \in \Ag $ is to almost-surely satisfy $ \gamma_i $. Now, consider a strategy profile $ \vec{\sigma} $ in which player 1/2 always chooses action $ a/b $ in $ s_0 $ ({\em i.e.}, chooses $ a/b $ with probability 1) 
%%	\mnnote{Not sure with probabilistic strategies\dots} \lhnote{This seems right to me, what are you not sure about?}.
%%	\mnnote{I guess my doubt was unfounded}
%	This is a Nash equilibrium since the goal of each player is achieved almost-surely, and thus they cannot benefit from changing their strategies. Furthermore, consider a strategy profile $ \vec{\sigma}' $ in which player 1/2 chooses action $ \bar{a}/\bar{b} $ with non-zero probability; this is also a Nash equilibrium, because unilateral deviation (either player 1 or player 2 chooses action $ a $ or action $ b $, respectively, with probability 1) cannot improve the situation: the path induced will still satisfy $ \neg \gamma_i $ with non-zero probability for each player $ i \in \Ag $ in the game.
Suppose we have a sender and a receiver who want to transmit some data. The sender and the receiver can be in either of two modes: idle or communicating. The data is sent sequentially in $ n $ blocks and successful transmission is assumed to be continuous, \textit{i.e.}, no gaps (missing blocks) are allowed between blocks; if there are gaps, the transmission fails and has to be restarted from the beginning. In order to be able to send the data, both sender and receiver have to be in the communicating mode. Furthermore, suppose that the network is noisy, thus for each block being transmitted, it may be lost with probability $ p $. 

To capture this, consider a game with $ \Ag = \{S, R\} $ representing the sender $S$ and receiver $R$. The set of actions for player $ j \in \Ag $ is $ \Ac_j = \{ c_j, i_j \} $, where $ c_j $ and $i_j$ mean that player $ j $ is communicating or idle,  respectively. The arena of the game is shown in Figure~\ref{fig:ex1ts}. Being in state $ s_i $ indicates that $ i $ blocks have been successfully transmitted, with $ s_n $ forming a sink state. The goal of each player $ j $ is to almost-surely satisfy $ \gamma_j = \sometime \psi $. 

There are infinitely many Nash equilibria in this game. However they can be classified into two categories: (a) those that satisfy the goals of each player; and (b) those that do not. For category (a), observe that any strategy profile that prescribes the action $ c_j $ for each player $ j $ with probability strictly greater than zero in every state $ s \in \{s_0,\dots,s_{n-1}\} $ is a Nash equilibrium. For (b), any strategy profile that prescribes $ i_j $ with probability one for each player $ j $ in any state $ s \in \{s_0,\dots,s_{n-1}\} $ is also a Nash equilibrium. 

Thus, the answer \textsc{E-Nash} with $ \varphi \equiv \sometime \psi $ (\textit{i.e.}, the data is eventually sent) is ``yes'', since there exist Nash equilibria that satisfy $ \varphi $ with probability one, namely, the strategy profiles that belong to category (a). On the other hand, the answer to the \textsc{A-Nash} query with the same property $\varphi$ is ``no'', since we have equilibria that belong to category (b).
\end{example}

% \begin{remark}
Similar, but more realistic, versions of Example~\ref{ex:1} can be constructed. For example, consider a distributed system with $n$ servers $\{S_i\}_{0 \leq i \leq n}$, each of which has a message inbox (a FIFO queue channel) that can contain up to $k$ messages. Each server $S_i$ can send a message $m$ to another server $S_j$, where  $m \in \Sigma$ for some finite $\Sigma$. This happens \emph{instantaneously} in that $m$ is placed on the FIFO channel of the receiver, although with a probability $p \in (0,1)$ that this fails. We denote this action by $snd(i,j,m)$. Each server can also opt to perform two other actions: pop the first message in the inbox (denoted by $pop(j,i,m)$, meaning that $S_j$ pops message $m$ sent by $S_i$), or remain idle. In addition, we require that each server must pop a message when the inbox is full. The system operates fully concurrently: each server acts completely independently of each other. 

The goal of each server $S_i$ is $$ \gamma_i = \bigwedge_{j,m} \always( snd(i,j,m) \to \sometime pop(j,i,m) );$$ namely, each message that is sent has to be eventually received. It is not difficult to show that given the property $ \phi = \bigwedge_{i=1}^{n} \gamma_i$, both \textsc{E-Nash} and \textsc{A-Nash} queries return ``yes'' answers. This is because a server $S_i$ that has sent a message $m$ to server $S_j$ can resend the message, until it knows that the message has been popped by $S_j$, thus forcing the \LTL goal to be satisfied almost surely. 
 
The above system is an example of a so-called stochastic lossy channel system \cite{SLCS-abdulla,SLCS-bertrand}, but restricted to bounded channels, which is reasonable in practice, wherein memories are bounded. Of course, communicating systems that employ channels are quite realistic in practice (as in, {\em e.g.}, the Erlang programming language), and handling the possibility of a message loss is important in the study of large communicating and distributed computer systems in general.
% \qedsymbol
% \end{remark}

In the remainder of the paper, it will be useful to sometimes consider a two-player zero-sum variant of an existing game in which the set of players is partitioned into two coalitions, $A \subseteq \Ag$ and $\Ag \setminus A$. In the case of \LTL objectives then $A$ has goal $\psi$ and $\Ag \setminus A$ has goal $\neg \psi$, and for parity objectives $A$ has even parity for priority function $ \alpha : \St \to \Nat $ and $\Ag \setminus A$ odd parity. We define this formally as follows. 

\begin{definition}
	\label{def:coalition_game}
	Let $\Game = (\CGModel, (\gamma_i)_{i \in \Ag}, \labFun)$ be a CSG whose underlying arena is $\CGModel = (\Ag, (\Ac_{i})_{i \in \Ag}, \St, s^{0}, \trnFun)$ and let $A \subseteq \Ag$. Then the \emph{two-player coalition game arena} is defined as 
	$\CGModel^{A} = ((1, 2), (\times_{i \in A}\Ac_{i}, \times_{i \in \Ag \setminus A}\Ac_{i}), \St, s^{0}, \trnFun_{A})$ where $\trnFun_{A}(s, (a_1,a_2)) = \trnFun(s, (\vec{a}_{A}, \vec{a}_{-A}))$. The \emph{two-player \LTL coalition game} with respect to $\Game$, $A$, and \LTL formula $\psi$, is thus defined as $\Game^{A,\psi} = (\CGModel^{A}, (\psi, \neg \psi), \labFun)$, and the \emph{two-player parity coalition game} with respect to $\Game$, $A$, and priority function $\alpha$, is defined as $\Game^{A,\alpha} = (\CGModel^{A}, (\alpha, \bar{\alpha}))$ where $\bar{\alpha}(s) = \alpha(s) + 1$ for any state $s \in \St$.
\end{definition}

In the remaining subsections we address the three main decision problems considered in the non-cooperative setting.
% \textsc{Membership}, \textsc{E-Nash}, and \textsc{A-Nash}.

\subsection{\textsc{Membership}}

Recall that the \textsc{Membership} problem requires two inputs: a game $ \Game $ and a strategy profile $ \vec{\sigma} $. We then ask if $ \vec{\sigma} $ forms a Nash equilibrium. Note that, in general, infinite memory strategies are needed to play concurrent $ \omega $-regular games with almost-sure winning conditions~\cite{CH2012}, however for this problem we assume that the input $ \vec{\sigma} $ is represented by some finite state transducer. An optimal procedure for solving \textsc{Membership} is given by Algorithm \ref{alg:membsership}, as shown by the following theorem.

\begin{theorem}
    \label{membership}
  \textsc{Membership}  is \twoexptime-complete.
\end{theorem}

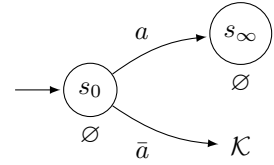
\begin{figure}[]
	\begin{center}
		\begin{tikzpicture}
		[->,>=latex,shorten >=1pt,auto,node distance=1cm, auto, sq node/.style={minimum size=0.5cm,draw}]
		
		\node[style={circle,draw},label=below:{$\varnothing$}] (0) [] {$ s_{0} $};
		\node[style={circle,draw},label=below:{$\varnothing$}] (1) [right of=0,xshift=1cm,yshift=0.75cm] {$ s_{\infty} $};
		\node[] (3) [right of=0,xshift=1cm,yshift=-0.75cm] {$ \MDP  $};
		
		\path
		++ (-1,0) edge node[]{} (0)
		(0) edge [bend right =-15] node[]{$ a $} (1)
		(0) edge [bend right =15] node[xshift=-0.5cm,yshift=-0.45cm]{$ \bar{a} $} (3)
		;
		
		\end{tikzpicture}
	\end{center}
	\caption{The CSGA for our reduction from \textsc{Membership} to qualitative \LTL model checking in MDPs. Edges without probability labels mean they have probability 1.}
	\label{fig:mem_reduction}
\end{figure}

\begin{proof}
Observe that checking line \ref{alg:membsership0} of Algorithm \ref{alg:membsership} amounts to (qualitative) model checking of the \LTL formula $ \gamma_i $ on the resulting MC $ \MC_{\vec{\sigma}} $, that is, after non-determinism in $ \CGModel $ is resolved by $ \vec{\sigma} $. This step can be done in \pspace~\cite{CY95}. Checking line \ref{alg:membsership1} amounts to \LTL model checking on the CSGA $ \CGModel_{\vec{\sigma}_{-i}} $, {\em i.e.}, model checking \LTL over MDP, which is \twoexptimeC~\cite{CY95}. Therefore, we have a \twoexptime procedure for solving \textsc{Membership}. 

For hardness, we reduce from qualitative \LTL model checking on MDPs. Given an MDP $ \MDP $ with labelling function $\lambda$ and an \LTL formula $ \phi $, we build a corresponding \textsc{Membership} instance $ (\Game,\vec{\sigma}) $ as follows. $ \Game = (\CGModel,\gamma_1,\lambda') $ is a CSG, where $ \gamma_1 = \ltlnext(\phi \wedge p) $ and $ p $ is a fresh variable. The CSGA $ \CGModel = (\Ag, \St, s^0, (\Ac_{i})_{i \in \Ag}, \trnFun) $ is built from $ \MDP  $ with only a single player, two additional states, and two fresh additional actions. Formally, $ \Ag = \{1\} $, $ \St = S \cup \{s_{\infty},s^0\} $, $ \Ac_1 = \Ac \cup \{ a,\bar{a} \} $, and
\[
\trnFun(s,a) =
\begin{cases}
s_{\infty}, & \text{if $(s,a_1) = (s^0,a)$ or $ s = s_{\infty} $} \\
s_{\iota}, & \text{if $(s,a_1) = (s^0,\bar{a})$} \\
\Pr(s,a), & \text{otherwise.	} \\
\end{cases}
\]
The labelling function $\lambda'$ is the same as $\lambda$ except that $ \lambda(s_{\infty}) = \varnothing $ and $\lambda'(s_\iota) = \lambda(s_\iota) \cup \{p\} $. An illustration of the construction of $ \CGModel $ is shown in Figure~\ref{fig:mem_reduction}. 

The strategy profile is defined as $ \vec{\sigma} = (\sigma_1) $, where $ \sigma_1(s^0) = a $, {\em i.e.}, action $ a $ is chosen with probability 1. Observe that a ``yes'' answer to the \textsc{Membership} query means that $ \vec{\sigma} \in \NE(\Game) $, which implies that $ \phi $ is not satisfied in $ \MDP $ with probability one. On the other hand, a ``no'' answer ({\em i.e.}, $ \vec{\sigma} \not\in \NE(\Game) $) implies that $ \phi $ is satisfied in $ \MDP $ with probability one. Furthermore, the construction can be done in polynomial time, concluding the proof.
\end{proof}

\begin{algorithm}[t]
	\caption{\textsc{Membership}}
\begin{algorithmic}[1]
	\Input $ \Game, \vec{\sigma} $
	\For{$ i \in \Ag $}
	\If{$ \vec{\sigma} \not \models \AS(\gamma_i) $ \label{alg:membsership0}}
		\If{ $ \exists \strElm_{i}' \in \Sigma_i$ s.t. $ (\vec{\sigma}_{-i}, \strElm_{i}') \models \AS(\gamma_i) $ \label{alg:membsership1}}
	\State \textbf{return} ``no''
	\EndIf
	\EndIf
	\EndFor
	\State \textbf{return} ``yes''
\end{algorithmic}
\label{alg:membsership}
\end{algorithm}

\subsection{\textsc{E-Nash} and \textsc{A-Nash}}

%\mnnote{Below is some note from last discussion. We are using ``almost-sure'' winning criterion here.}

For a given $ \Game $ and formula $ \phi $, \textsc{E-Nash} asks whether \textit{some} Nash equilibrium almost-surely satisfies $ \phi $. On the other hand, \textsc{A-Nash} asks whether \textit{all} Nash equilibria almost-surely satisfy $ \phi $. 
Observe that \textsc{A-Nash} is closely related to \textsc{E-Nash}, \textit{i.e.}, it can be framed as an instance of \textsc{E-Nash}, with a small modification: rather than check whether the formula $ \AS(\phi) $ is satisfied in some Nash equilbrium, we check if $ \NZ(\neg \phi) $ is satisfied. A Nash equilbirum that satisfies $ \NZ(\neg \phi) $ is a negative witness to \textsc{A-Nash}. Thus, we first provide a decision procedure for solving \textsc{E-Nash}, and later adapt the procedure to handle the  \textsc{A-Nash} problem.
%\lhnote{This isn't true, $\AS(\neg\varphi) \not\equiv \neg \AS(\varphi)$, cf. the section on \textsc{E-Core}.}

To solve the problem, we adapt the technique presented in \cite{GutierrezNPW20}. At this point, it is important to note that while our approach is inspired by the one proposed in \cite{GutierrezNPW20}, the setting considered in this paper differs in multiple ways. Firstly, here we consider stochastic games and randomised strategies, while \cite{GutierrezNPW20} only considers deterministic games and pure strategies. Secondly, we allow strategies to have infinite number of states, instead of finite (albeit unbounded) states. Finally, we use almost-sure winning conditions, which does not apply to deterministic games played with pure strategies.
%Thus, our approach here is more general compared to the one in the aforementioned paper.

To describe our approach, we begin with some definitions.
%
%\begin{theorem}[\cite{Zielonka2004PerfectInformationSP,Chatterjee2004QuantitativeSP}]
%	For turn-based stochastic ($ 2\frac{1}{2} $-player) parity games with a finite state space, both players have optimal pure positional strategies.
%\end{theorem}
%
%\lhnote{This result is only for turn-based parity games though, right? And don't we need concurrent ones (CSPGs), in which winning strategies may require infinite memory?}
%
%The result above is important since it allows us to reason about punishing region locally \cite{GutierrezNPW20}. To carry out such a reasoning, we transform a CSG $ \Game $ into its associated CSPG $ \ParGame $. The basic idea is to transform each \LTL goal $ \gamma_i $ into its corresponding DPW $ \Automaton_{\gamma_{i}} $. Using the collection of DPWs $ (\Automaton_{\gamma_i})_{i \in \Ag} $, we build the underlying arena of $ \ParGame $.
%
%\begin{definition}
%	\label{def:ltltopar}
	Let $\Game = (\CGModel, \labFun, (\gamma_i)_{i \in \Ag})$ be a CSG whose underlying arena is 
	$\CGModel = (\Ag, (\Ac_{i})_{i \in \Ag}, \St, s^{0}, \trnFun)$, and 
	%Moreover, 
	let $\Automaton[\gamma_{i}] = \tuple{\pow{\AP}, Q_{i}, q_{i}^{0}, \rho_{i}, \alpha_{i}}$ be the DPW corresponding to player $i$'s goal $\gamma_{i}$ in $\Game$, and $ \Automaton[\phi]  = \tuple{\pow{\AP}, Q_{\phi}, q_{\phi}^{0}, \rho_{\phi}, \alpha_{\phi}} $ to the formula $ \phi $.
	The {\em CSPG~$\ParGame$ associated to~$\Game$}~is $\ParGame = (\CGModel', (\alpha_i')_{i \in \Ag},\alpha_{\phi}')$, where $\CGModel' = (\Ag, (\Ac_{i})_{i \in \Ag}, \St', s^{0\prime}, \trnFun[]['])$ and $(\alpha_i')_{i\in\Ag}$ are as follows:
	
	\begin{itemize}
		\item
		$\St' = \St \times \bigtimes_{i \in \Ag} Q_{i} \times Q_{\phi}$ and $s^{0\prime} = (s^{0}, q^{0}_{1}, \ldots , q^{0}_{n},q^{0}_{\phi})$;
		
		%		\item
		%		$s_{0}' = (s_{0}, q^{0}_{1}, \ldots , q^{0}_{n})$;
		
		\item
		for each state $(s, q_{1}, \ldots , q_{n},q_{\phi})\in \St'$ and action profile $\vec{a}$, we define $\trnFun'((s, q_{1}, \ldots , q_{n},q_{\phi}), \vec{a}) = (\trnFun(s, \vec{a}), \rho_{1}(q_{1}, \labFun(s)), \ldots, \rho_{n}(q_{n}, \labFun(s)), \rho_{\phi}(q_{\phi},\labFun(s)))$;			
		
		\item $\alpha_i'(s, q_{1}, \ldots q_{n},q_{\phi}) = \alpha_{i}(q_{i})$.
		\item $ \alpha_{\phi}'(s, q_{1}, \ldots q_{n},q_{\phi}) = \alpha_{\phi}(q_{\phi}) $
	\end{itemize}
	
%\end{definition}

Observe that in the translation of $ \Game $ to its associated $ \ParGame $, the set of actions for each player is unchanged. Therefore, the set of strategies in both $ \Game $ and $ \ParGame $ is the same, since for every state $ s \in \St $ and action profile $ \vec{a} $, it follows that $ \vec{a} $ is available in $ s $ if and only if it is available in $ (s,q_1,\dots,q_n,q_{\phi}) \in \St' $, for all $ (q_1,\dots,q_n,q_{\phi}) \in \bigtimes_{i \in \Ag} Q_i \times Q_{\phi} $. 
%Furthermore, the construction of $ \trnFun' $ preserves probability distribution assignment induced by $ \trnFun $. 
This, in turn, means that, for a given strategy profile $ \vec{\sigma} $, we obtain MCs $ \MC_{\vec{\sigma}} $ and $ \MC_{\vec{\sigma}}^{\Par} $ that correspond to $ \Game $ and $ \ParGame $, respectively. Furthermore, 
%let $ \paths(\MC_{\vec{\sigma}}^{\Par}, s) = \{ \pi = (s_0)_1(s_1)_1\dots \in \St^{\omega} : s_0 = s, \forall k \in \Nat, \trnFun(s_k,s_{k+1}) > 0 \} $ 
since the construction of $ \trnFun' $ preserves the probability distribution assignments induced by $ \trnFun $, we have the following lemma.

\begin{lemma}\label{lemma:prob-inv}
	For a CSG $ \Game $ and its associated CSPG $ \ParGame $, it holds that $ \pi' \in \paths(\MC_{\vec{\sigma}}^{\Par},s^{0\prime}) $ if and only if $ \proj_s(\pi') \in \paths(\MC_{\vec{\sigma}},s^0) $, where $ \proj_s(\pi') $ is the $ s $ component of $ \pi' $.
\end{lemma}

 Now, suppose that $ \vec{\sigma} \models \AS(\gamma_{i}) $ in $ \Game $; thus we have that $ \Pr_{\MC_{\vec{\sigma}}}(\{ \pi \in \paths(\MC_{\vec{\sigma}},s^0) : \pi \models \gamma_{i} \}) = 1 $. Moreover, consider the component $ \proj_{q_i}(\pi'), i \in \Ag $. By the construction of $ \ParGame $, it holds that $ \proj_{q_i}(\pi') $ is the run executed by the DPW $ \Automaton_{\gamma_{i}} $ when $ \lambda(\pi) $ is read, and the parity of $ \pi' $ with respect to $ \alpha_i' $ corresponds to the one recognised by $ \Automaton_{\gamma_{i}} $. Thus, by Lemma~\ref{lemma:prob-inv}, it holds that $ \Pr_{\MC_{\vec{\sigma}}^{\Par}}(\{ \pi' \in \paths(\MC_{\vec{\sigma}}^{\Par},s^{0\prime}) : \proj_{q_i}(\pi') \models \alpha_{i}' \}) = 1 $, which implies that $ \vec{\sigma} \models \AS(\alpha_{i}) $ in $ \ParGame $. Therefore, we obtain the following lemma.

\begin{lemma}
	\label{lem:neinv}
	For a CSG $ \Game $ and its associated CSPG $ \ParGame $, it is the case that for every strategy profile $ \vec{\sigma} $ and player $ i $, $ \vec{\sigma} \models \AS(\gamma_i) $ if and only if $ \vec{\sigma} \models \AS(\alpha_{i}) $.
\end{lemma}

With Lemma~\ref{lem:neinv} in hand, we can show that the set of Nash equilibria for any CSG $ \Game $ exactly corresponds to the set of Nash equilibria of its associated CSPG $ \ParGame $. Formally, we have the following proposition.

\begin{proposition}
    \label{thm:ne-equiv}
	Given a CSG $ \Game $ and its associated CSPG $ \ParGame $, we have $ \NE(\Game) = \NE(\ParGame) $.
\end{proposition}

\begin{proof}
	We prove the proposition by double inclusion. Assume $ \vec{\sigma} \in \NE(\Game)$, and, by contradiction, $ \vec{\sigma} \not\in \NE(\ParGame) $. Due to Lemma~\ref{lem:neinv}, it holds that $ W_{\Game}(\vec{\sigma}) = W_{\ParGame}(\vec{\sigma}) $. Then, there is a player $ j \in L_{\Game}(\vec{\sigma}) $ and a strategy $ \sigma_j' $ such that $ (\vec{\sigma}_{-j,},\sigma_j') \models \AS(\alpha_j) $ in $ \ParGame $. This implies that $ \sigma_j' $ is also a beneficial deviation for $ j $ in $ \Game $ -- a contradiction. On the other hand, for every $ \vec{\sigma} \in \NE(\ParGame) $, we can also reason in a symmetric way to conclude that $ \vec{\sigma} \in \NE(\Game) $.
\end{proof}

Proposition~\ref{thm:ne-equiv} allows us to compute Nash equilibria in CSG $ \Game $ via its associated CSPG $ \ParGame $. To do this, we use Nash equilibrium characterisation presented in \cite{GHW15} which employs two concepts: \textit{punishment} and \textit{attributability}. For the former, we introduce the notion of \textit{punishing strategy}.

\begin{definition}
	For a game $ \ParGame $, player~$ j $, and state $ s $, the strategy profile $ \vec{\sigma}_{-j} $ is a \textit{punishing strategy} for player~$ j $ in $ s $ if $ ((\vec{\sigma}_{-j},\sigma_j'),s) \models \neg \AS(\alpha_j) $, for every possible $ \sigma_j' $.
\end{definition}

We say that a state $s$ is punishing for $j$ if there exists a punishing strategy profile for $j$ on $s$.
Moreover, we denote by $\Pun_{j}(\ParGame)$ the set of punishing states for player $ j $ in $\ParGame$.
To compute $\Pun_{j}(\ParGame)$, we solve the two-player parity coalition game $ \ParGame^{A,\alpha_j} $ where $ A = \Ag \setminus \{j\} $. Let $ \WinSet_{A}(\ParGame^{A,\alpha_j}) $ be the set of winning states of coalition player $ A $ in $ \ParGame^{A,\alpha_j} $. Then the set $ \WinSet_{A}(\ParGame^{A,\alpha_j}) $, corresponds exactly to $ \Pun_j(\ParGame) $.
A pair $(s, \jact) \in \St \times \AcProf$ is \emph{punishing-secure} for player $j$, if $\spt(\trnFun(s, (\jact_{-j},\act'_j))) \subseteq \Pun_{j}(\ParGame)$ for every action $\act_j'$. 
We can then extend the notion of punishing-secure pairs just defined to MCs as follows.

\begin{definition}
	Given a CSPG $ \ParGame $ and strategy profile $ \vec{\sigma} $, the associated MC $ \MC_{\vec{\sigma}} = (S,s_\iota,\trnFun',\lambda) $ is punishing-secure for $ j \in L_{\ParGame}(\vec{\sigma}) $ if for every $ s, s' \in S $ and every associated $ \vec{a} \in \AcProf $ such that $ \trnFun'(s,s') > 0 $, $ (s,\vec{a}) $ is punishing-secure.
\end{definition}

Now, with those definitions in place, we can characterise Nash equilibria in CSPGs as follows.

\begin{proposition}
    \label{thm:ne-char}
%	Let $ \vec{\sigma} $ be a strategy profile, $ \MC_{\vec{\sigma}} = (S,s_\iota,\trnFun',\lambda) $ be the associated MC of a given CSPG $ \ParGame $.
	Let $ \MC_{\vec{\sigma}} $ be the associated MC of a given CSPG $ \ParGame $ and strategy profile $ \vec{\sigma} $.
	It holds that $ \vec{\sigma} \in \NE(\ParGame) $ if and only if for every player $ j \in L_{\ParGame}(\vec{\sigma}) $, $ \MC_{\vec{\sigma}} $ is punishing-secure for $ j $.
%	In a given CSPG $ \ParGame $, there exists a Nash equilibrium if and only if there exists a strategy profile $ \vec{\sigma} $ such that, $ \vec{\sigma} \models \bigwedge_{i \in W_{\ParGame}(\vec{\sigma})} \AS(\gamma_{i}) $ and for all $ s, s' \in S$ of the associated MC $ \MC_{\vec{\sigma}} = (S,s_\iota,\trnFun',\lambda) $, such that $ \trnFun'(s,s') > 0 $
\end{proposition}

\begin{proof}
	From left to right, suppose $ \vec{\sigma} \in \NE(\ParGame) $, and assume for a contradiction that $ \MC_{\vec{\sigma}} = (S,s_\iota,\trnFun',\lambda) $ is not punishing-secure for some $ j \in L_{\ParGame}(\vec{\sigma}) $. This means that there is a state $ s \in S $ and action $ a_j' \in \Ac_j $ such that there exists $ s' \in \spt(\trnFun(s,(\jact_{-j},a_j')))$ where $s' \not\in \Pun_{j}(\ParGame) $. This, in turn, means that there exists a (deviating) strategy of player $ j $ such that there is non-zero probability of player $ j $ escaping the punishing area. By the determinacy of two-player concurrent parity games with almost-sure winning conditions \cite{deAlfaro2000}, player $ j $ can thus achieve its goal with probability 1. Since this is a beneficial deviation for player $ j $, then $ \vec{\sigma} $ is not in the set of Nash equilibria of $ \ParGame $ -- which is a contradiction. 
	
	From right to left, we first assume the existence of some $ \MC_{\vec{\sigma}} $ that is punishing-secure for every losing player $ j $. Such an MC can be generated by a (possibly infinite state) transducer $ T $. Moreover, for every losing player $ j $ and every state $ s \in S $ of $ \MC_{\vec{\sigma}} $, there is a punishing strategy for $ j $. Combining $ T $ with such punishing strategies, we obtain a strategy profile $ \vec{\sigma} $ that follows $ T $, until a losing player $ j $ deviates. At this point, the concept of attributability is required, since to be able to punish, the coalition $ \Ag \setminus \{j\} $ needs to know who should be punished once a deviation happens. In order to do this, the players must be able to remember the history of play from the beginning of the game up until the point at which a deviation happens. Notice that in general this requires infinite memory. In such a case, $ \vec{\sigma} $ would start punishing player $ j $. Therefore, there is no beneficial deviation for player $ j $, and strategy profile $ \vec{\sigma} $ is a Nash equilibrium of the game.
\end{proof}

%Using the Nash equilbrium characterisation in Theorem~\ref{thm:ne-char}, we present an algorithm for \textsc{E-Nash} as follows.
Proposition~\ref{thm:ne-char} characterises Nash equilibria through the concept of a \textit{punishing region}, {\em i.e.}, a region where, for a given set of losing players $ L $, each player $ j \in L $ can be punished. This region, denoted as $ \ParGame^{-L} $, is the game resulting from $ \ParGame $ after the removal of the states that are not punishing for some $ j \in L $, and the edges $ (s,\vec{a}) $ that are not punishing-secure for some $ j \in L $. We next observe that a positive answer for Algorithm~\ref{alg:enash} corresponds to the existence of Nash equilibrium in $ \Game $ that satisfies $ \phi $ with probability one, and prove that it is optimal. 

\begin{theorem}\label{thm:enash}
	\textsc{E-Nash} and \textsc{A-Nash}  are \twoexptime-complete.
\end{theorem}

%A deterministic algorithm obtained from the procedure above is shown in Algorithm~\ref{alg:enash}.

\begin{proof}
Algorithm \ref{alg:enash} runs in doubly exponential time. The underlying structure $ \CGModel' $ of $ \ParGame $ is doubly exponential in the size of the \LTL goals of $ \Game $ and formula $ \phi $, but the priority functions sets $ (\alpha_i)_{i \in \Ag} $ and $ \alpha_{\phi} $ are only singly exponential \cite{Piterman2007}. Computing $ \Pun_j(\ParGame) $ is polynomial in the size of $ \CGModel' $ and exponential time in the size of priority functions set \cite{deAlfaro2000}. Line \ref{alg:enash0} in the algorithm corresponds to checking the realisability problem for a \textit{qualitative parity logic} formula containing conjunctions of almost-sure atoms over the MDP $\MDP^{-L}$ resulting from $\ParGame^{-L}$ when all players are as one, which can be solved in polynomial time \cite{BGR2020}. The formula expresses that the objective $ \alpha_{\phi} $ (representing $ \phi $) is satisfied with probability 1, and each winning player $ i $ (that cannot be punished in $ \CGModel^{-L} $) achieves its goal with probability 1. The overall complexity of the algorithm is thus in \twoexptime.

Now that we have a procedure to solve \textsc{E-Nash}, we can adapt it to solve \textsc{A-Nash}. The adaptation is straightforward, and goes as follows. First, when building $ \ParGame $ from $ \Game $, instead of DPW $ \Automaton_{\phi} $, we use $ \Automaton_{\neg \phi} $, {\em i.e.}, a DPW built from $ \neg \phi $. Thus, in Algorithm~\ref{alg:enash}, the formula in line \ref{alg:enash0} is replaced by $\NZ(\alpha_{\neg \phi}) \wedge \bigwedge_{i \in W} \mthfun{AS}(\alpha_i)$ and the positive and negative answers in lines \ref{alg:enash1} and \ref{alg:enash2} respectively are swapped. Thus we also have a \twoexptime algorithm for \textsc{A-Nash}. 
%\lhnote{See my earlier comment about this.}

For hardness, we reduce from qualitative \LTL model checking over MDPs. Given an MDP $\MDP$ with a labelling function $\lambda$ and formula $\varphi$, then solving the \textsc{E-Nash} problem with input given by the one-player game $\Game = (\MDP, (\top), \lambda)$ and formula is equivalent to \LTL model checking over the MDP. This fact, and the duality between \textsc{A-Nash} and \textsc{E-Nash} concludes the proof.
\end{proof}

\begin{algorithm}[t]
	\caption{\textsc{E-Nash}}
	\label{alg:enash}
	\begin{algorithmic}[1]
		
		%		\Procedure{Euclid}{$a,b$}\Comment{The g.c.d. of a and b}
		\Input $ \Game, \phi $
		
		\State build $ \ParGame $ from $ \Game $
		\For{$ W \subseteq \Ag $}
		\For{$ j \in L = \Ag \setminus W $}
		\State compute $ \Pun_j(\ParGame) $
		\EndFor
		
		\State build $ \ParGame^{-L} $ and obtain $ \MDP^{-L} $
		
		\If{$ \MDP^{-L} \models \AS(\alpha_{\phi}) \wedge \bigwedge_{i \in W} \mthfun{AS}(\alpha_i) $ \label{alg:enash0}}
		\State \textbf{return} ``yes'' \label{alg:enash1}
		\EndIf
		\EndFor
		
		\State \textbf{return} ``no'' \label{alg:enash2}
		%		\EndProcedure
	\end{algorithmic}
\end{algorithm}

We now turn our attention to the cooperative setting, in which equilibria are instead characterised by the \emph{core}. 

%\newpage 
\section{Cooperative Rational Verification}
\label{secn:coop}

Nash equilibrium is a non-cooperative solution concept: it assumes that players must act in isolation, without the possibility of forming binding agreements to cooperate. In many settings, however, binding agreements are possible, and for these it is appropriate to consider \emph{cooperative} solution concepts, of which the \emph{core} is the most prominent. While Nash equilibrium considers strategy profiles that are stable against individual deviations, the core considers possible beneficial deviations by groups of players (\emph{coalitions}). 

We follow the definition of \textit{core} from \cite{GKW2019}. 
%
%For a given game $ \Game $ and a strategy profile $ \vec{\sigma} $, we write $ W(\vec{\sigma}) $ to denote the set of players that would get their goal achieved (the ``winners'') if the strategy profile $ \vec{\sigma} $ is used, and $ L(\vec{\sigma}) $ to denote the set of players that not:
%
%\begin{align*}
%	W(\vec{\sigma}) &= \{ i \in \Ag : \vec{\sigma} \models \gamma_i \}\\
%	L(\vec{\sigma}) &= \Ag \setminus W(\vec{\sigma}).
%\end{align*}
%
We first define the notion of a \textit{deviation} and a \textit{beneficial deviation}. A deviation is a joint strategy $ \vec{\sigma}_{A} $ for the coalition $ A \subseteq \Ag $, with $ A \neq \varnothing $. For a strategy profile $ \vec{\sigma} $, we say $ \vec{\sigma}_{A}' $ is a beneficial deviation from $ \vec{\sigma} $ if $ A \subseteq L(\vec{\sigma}) $ and for all $ \vec{\sigma}_{-A}' $, we have $ A \subseteq W((\vec{\sigma}_{A}',\vec{\sigma}_{-A}')) $.
% \begin{itemize}
%     \item $ A \subseteq L(\vec{\sigma}) $
% 	\item For all $ \vec{\sigma}_{-A}' $, we have $ A \subseteq W((\vec{\sigma}_{A}',\vec{\sigma}_{-A}')) $.
% \end{itemize}
The core of a game $ \Game $, denoted $ \coreset(\Game) $, is then defined to be the set of strategy profiles that admit no beneficial deviation. 

Given the above definitions, we can introduce the key decision problems relating to rational verification and the core.

\begin{quote}
	\underline{\textsc{E-Core}}:\\
	\emph{Given}: Game $\Game$, \LTL formula $\phi$.\\
	\emph{Question}: Is it the case that
	{\flushright $\qquad\qquad\quad  \exists\vec{\sigma}\in\coreset(\Game).\ \vec{\sigma}\models \AS(\phi)$?}
\end{quote}
	
\begin{quote}
	\underline{\textsc{A-Core}}:\\
	\emph{Given}: Game $\Game$, \LTL formula $\phi$.\\
	\emph{Question}: Is it the case that
	{\flushright $\qquad\qquad\quad 
	\forall\vec{\sigma}\in\coreset(\Game).\ \vec{\sigma}\models \AS(\phi)$?}
\end{quote}

\begin{example}\label{ex:2}
	Recall the game in Figure \ref{fig:ex1ts}. As we saw in Example \ref{ex:1}, the set of strategy profiles $ \vec{\sigma} $ in which player S/R chooses action $ i_S/i_R $ with probability one in any $ s \in \{ s_0,\dots,s_{n-1} \} $ is a Nash equilibrium, because unilateral deviations cannot improve the situation. However, this strategy profile is \emph{not} in the core, because there is a cooperative beneficial deviation to the strategy $ \vec{\sigma} $ in which player S/R chooses action $ c_S/c_R $ with probability strictly greater than zero in every state $ s \in \{ s_0,\dots,s_{n-1} \} $. This means that, while the \textsc{A-Nash} query with the property $ \varphi \equiv \sometime \psi $ returns ``no'', the \textsc{A-Core} query with the same property returns ``yes'', since every strategy profile in the core satisfies $ \varphi $ with probability one -- in general they are always Pareto-optimal.
\end{example}

Alongside \textsc{E-Core} and \textsc{A-Core}, we will also often be interested in the question of whether a particular alternative strategy $\vec{\sigma}'_{A}$ represents a beneficial deviation from $\vec{\sigma}$ for a coalition $A$ of players. This question, along with a version of \textsc{Membership} for cooperative games, forms the final two decision problems we investigate in this work.

\begin{quote}
	\underline{\textsc{Core Membership}}:\\
	\emph{Given}: Game $\Game$, strategy profile $\vec{\sigma}$.\\
	\emph{Question}: Is it the case that $\vec{\sigma} \in \coreset(\Game)$?
	\end{quote}
	
\begin{quote}
	\underline{\textsc{Beneficial Deviation}}:\\
	\emph{Given}: Game $\Game$, strategy profile $\vec{\sigma}$, deviation $\vec{\sigma}'_{A}$.\\
	\emph{Question}: Is $\vec{\sigma}'_{A}$ a beneficial deviation from $\vec{\sigma}$ in $\Game$?
\end{quote}
As noted above, the core can be viewed as the set of strategy profiles from which no \emph{coalition} (as opposed to \emph{player} in the case of NE) has a beneficial deviation. Our complexity results for this cooperative solution concept follow a similar high-level line of reasoning as taken in previous work in the non-stochastic setting \cite{GKW2019}. However, as it will be seen next, the ``inner workings'' to obtain the main complexity results rely on very different techniques needed to be able to account for the various probabilistic features in the game.

\subsection{\textsc{E-Core} and \textsc{A-Core}}

We begin by noting that the ability of a coalition $A$ to achieve an \LTL goal $\psi$ can be interpreted as its possession of a winning strategy in the two-player coalition game $\Game^{A,\psi}$, as defined in the previous section. We say that such a game is \emph{winnable} if player 1 has a strategy for achieving $\psi$. Using these concepts, we restate the following result, with some adaptations for our stochastic setting.

\begin{lemma}[\cite{GKW2019}] 
	Let $\Game = (\CGModel, \labFun, (\gamma_i)_{i \in \Ag})$ be a CSG whose underlying arena is 
	$\CGModel = (\Ag, (\Ac_{i})_{i \in \Ag}, \St, s^{0}, \trnFun)$ and let $\varphi$ be an \LTL formula. Then $\Game$ and $\varphi$ satisfy \textsc{E-Core} if and only if there exists $W \subseteq \Ag$ such that
	\begin{itemize}
		\item There exists some $\vec{\sigma}$ such that $\vec{\sigma} \models \chi_W$
		\item For all $L \subseteq \Ag \setminus W$, $\Game^{L,\psi_L}$ is not winnable
	\end{itemize}
	where $\chi_W = \AS(\varphi) \wedge \bigwedge_{i \in W} \AS(\gamma_i) \wedge \bigwedge_{i \in \Ag \setminus W} \neg \AS(\gamma_i)$ and $\psi_L = \bigwedge_{i \in L} \AS(\gamma_i)$.
\end{lemma}

This result leads us to the procedure shown in Algorithm \ref{alg:e-core} for determining whether some game $\Game$ and \LTL formula $\varphi$ satisfy \textsc{E-Core}. Moreover, due to the duality between $\AS$ and $\NZ$ winning conditions we may express the \textsc{A-Core} problem for a game $\Game$ and formula $\varphi$ as the negation of the \textsc{E-Core} problem for $\Game$ where the conjunct $\AS(\varphi)$ is replaced by $\neg \AS(\varphi) \equiv \NZ(\neg \varphi)$. We thus have the following complexity results.

\begin{theorem}
	\textsc{E-Core} and \textsc{A-Core} are \twoexptimeC.
\end{theorem}

\begin{proof}
	The loop in line \ref{alg:e-core0} and the check in line \ref{alg:e-core3} (for each $A \subseteq N$) in Algorithm \ref{alg:e-core} are executed $2^{\vert\Ag\vert}$ times each. Next, observe that we may also write $\chi_W$ as $\AS(\varphi) \wedge \bigwedge_{i \in W} \AS(\gamma_i) \wedge \bigwedge_{i \in \Ag \setminus W} \NZ(\neg\gamma_i)$, a conjunction of $\AS$ and $\NZ$ conditions. Thus, by expressing $\varphi$, $\gamma_i$ for $i \in W$, and $\neg \gamma_i$ for $i \in \Ag \setminus W$ as DPWs and constructing the game $\ParGame$, we may perform this model checking problem in time polynomial in the size of $\ParGame$ using qualitative parity logic \cite{BGR2020}. 
    % As a DPW formed from an \LTL formula $\psi$ has $O(2^{2^{\vert \psi_A \vert}})$ states and $O(2^{\vert \psi_A \vert})$ labels \cite{Piterman2007}, then we see that $\ParGame \times \Automaton_\varphi$ has size doubly exponential in the original input and hence that the model checking step is in \twoexptime.
    As a DPW formed from an \LTL formula $\psi$ has states and labels doubly and singly exponential in the size of $\psi$ respectively \cite{Piterman2007}, then we see that $\ParGame$ has size doubly exponential in the original input and hence that the model checking step in the algorithm can be solved in \twoexptime.

	For line \ref{alg:e-core2} we begin by noting that $\vec{\sigma} \models \bigwedge_{i \in A} \AS(\gamma_i)$ if and only if $\vec{\sigma} \models \AS(\bigwedge_{i \in A} \gamma_i)$. Let us define $\psi^\wedge_A = \AS(\bigwedge_{i \in A} \gamma_i)$. Thus, we instead form the two-player parity coalition game $\ParGame^{A,\psi^\wedge_A}$. As $\vert \psi^\wedge_A \vert$ is linear in $\vert \gamma_1 \vert, \ldots, \vert \gamma_n \vert$, then the number of states in $\ParGame^{A,\psi^\wedge_A}$ is doubly exponential in the size of the original input, and the number of pairs in the parity accepting condition of $\ParGame^{A,\psi^\wedge_A}$ is singly exponential in the size of the original input. Whether player 1 has a winning strategy in this product game can be checked in time polynomial in the former and singly exponential in the latter for \AS-winning conditions \cite{deAlfaro2000}, meaning this step remains in \twoexptime as well.

	Thus, Algorithm \ref{alg:e-core} above can be seen to run in \twoexptime, providing an upper bound for \textsc{E-Core}. To see that this bound is tight, note that we may reduce (qualitative) model checking of \LTL over MDPs to \textsc{E-Core}. Given an MDP $\MDP$ with a labelling function $\lambda$ and formula $\varphi$, we input the one-player game $\Game = (\MDP, (\top), \lambda)$ and $\varphi$ to Algorithm \ref{alg:e-core}, which returns ``yes'' if and only if there is a strategy in $\MDP$ satisfying $\varphi$ almost-surely. As qualitative \LTL model checking over MDPs is in \twoexptime \cite{CY95}, and our construction is polynomial in the size of the original input, this concludes the proof.
\end{proof}

\begin{algorithm}[t]
\caption{\textsc{E-Core}}
\begin{algorithmic}[1]
	\Input $\Game$, $\varphi$
	\For{$A \subseteq \Ag$} \label{alg:e-core0}
		\State check if $\exists \vec{\sigma}$ s.t. $\vec{\sigma} \models \chi_A$, written $W_A$ \label{alg:e-core1}
		\State check that $\Game^{A,\psi_A}$ is not winnable, written $L_A$	\label{alg:e-core2}
	\EndFor
	\If{$\exists A \subseteq \Ag$ s.t. $W_A$ and $L_B$ $\forall B \subseteq \Ag \setminus A$} \label{alg:e-core3}
		\State \textbf{return} ``yes''
	\EndIf
	\Return ``no''
\end{algorithmic}
\label{alg:e-core}
\end{algorithm}

\subsection{\textsc{Core Membership}}

For the problem of \textsc{Core Membership} and \textsc{Beneficial Deviation} we assume, as in the non-cooperative setting, that the given strategies are all finite memory. By taking a product of the finite state transducers representing $\vec{\sigma}$ with the game $\Game$ then we may check \textsc{Core Membership} by first checking for every player $i$ whether $\vec{\sigma} \models \AS(\gamma_i)$, and then checking whether any subset $L \subseteq \Ag \setminus W$ can deviate to achieve the formula $\psi_L = \bigwedge_{i \in L} \AS(\gamma_i)$. This procedure is shown in full in Algorithm \ref{alg:core_mem} and gives rise to the following complexity result.

\begin{theorem}
	\textsc{Core Membership}  is \twoexptimeC.
\end{theorem}

\begin{proof}
	The first step (line \ref{alg:core_mem1}) of Algorithm \ref{alg:core_mem} is equivalent to performing qualitative \LTL model checking on the MC $ \MC_{\vec{\sigma}} $, which can be done in \pspace \cite{CY95}. The second step (line \ref{alg:core_mem1}) can be done in \twoexptime, as noted above, and thus we have that \textsc{Core Membership} is also in \twoexptime. As a lower bound we note that the problem of \textsc{Core Membership} is the same as the problem of \textsc{Membership} when $\vert N \vert = 1$, and so we may use exactly the same reduction given in the proof of Theorem \ref{membership} with the construction illustrated in Figure \ref{fig:mem_reduction}.
% 	where $\MC^{\neg\varphi}$ is a minimal (deterministic) MC satisfying $\neg \varphi$. We conclude the proof by observing that that there is a strategy in $\MDP$ that achieves $\varphi$ almost-surely if and only if Algorithm \ref{alg:core_mem} returns ``no'' with input $\Game = (\CGModel, (\ltlnext \varphi), \lambda)$ and $\vec{\sigma}$, where $\lambda$ is the combination of the labelling functions for $\MDP$ and $\MC^{\neg\varphi}$ with $\lambda(s_0) = \varnothing$ and $\vec{\sigma}$ is any strategy of size polynomial in the original input such that $\sigma_1(s_0) = a$.
\end{proof}

\begin{algorithm}[t]
	\caption{\textsc{Core Membership}}
	\begin{algorithmic}[1]
		\Input $\Game$, $\vec{\sigma}$
		\For{$i \in \Ag$}
			\State check if $\vec{\sigma} \models \AS(\gamma_i)$, written $W_i$ \label{alg:core_mem0}
		\EndFor
		\For{$L \subseteq \Ag \setminus \{i : W_i = \top\}$}
			\If{$\Game^{L,\psi_L}$ is winnable} \label{alg:core_mem1}
				\State \textbf{return} ``no''
			\EndIf
		\EndFor
		\Return ``yes''
	\end{algorithmic}
	\label{alg:core_mem}
\end{algorithm}

\subsection{\textsc{Beneficial Deviation}}

We now study \textsc{Beneficial Deviation}, solved using Algorithm \ref{alg:ben_dev}, and note that although the complexities for the other problems in cooperative rational verification are the same as in the deterministic setting, \textsc{Beneficial Deviation} is only in \pspace for deterministic games, and hence the stochastic setting is significantly more difficult.

\begin{theorem}
	\textsc{Beneficial Deviation}  is \twoexptimeC.
\end{theorem}

\begin{figure}[]
	\begin{center}
		\begin{tikzpicture}
		[->,>=latex,shorten >=1pt,auto,node distance=1cm, auto, sq node/.style={minimum size=0.5cm,draw}]
		
		\node[style={circle,draw},label=below:{$\varnothing$}] (0) [] {$ s_0 $};
		\node[] (1) [right of=0,xshift=1cm,yshift=1cm] {$ \MC^\varphi $};
		\node[] (2) [right of=0,xshift=1cm,] {$ \MC^{\neg\varphi} $};
		\node[] (3) [right of=0,xshift=1cm,yshift=-1cm] {$ \MDP  $};
		
		\path
		++ (-1,0) edge node[]{} (0)
		(0) edge [bend right =-25] node[]{$ a,b $} (1)
		(0) edge [] node[above]{$ a,\bar{b} $} (2)
		(0) edge [bend right =25] node[align=center,xshift=-0.75cm,yshift=-1cm]{$ \bar{a},b $\\$ \bar{a},\bar{b} $} (3)
		;
		
		\end{tikzpicture}
	\end{center}
	\caption{The CSGA for our reduction from \textsc{Beneficial Deviation} to qualitative \LTL model checking in MDPs. Edges without probability labels mean they have probability 1.}
	\label{fig:ben_dev_reduction}
\end{figure}
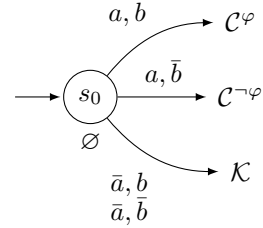

\begin{proof}
	We begin, in Algorithm \ref{alg:ben_dev}, by checking whether $\vec{\sigma} \models \bigwedge_{i \in A} \neg \AS(\gamma_i)$, or equivalently whether $\vec{\sigma} \models \NZ(\neg \gamma_i)$ for each player $i \in A$, as otherwise there is no beneficial deviation for the coalition $A$ and we are done (line \ref{alg:ben_dev0}). This is again a simple qualitative \LTL model checking problem on $ \MC_{\vec{\sigma}} $, which is in \pspace, as remarked in the previous proof. The second condition we check is whether, when the coalition $A$ is instead playing $ \vec{\sigma}_{A}' $, there exists a way for the remaining players $\Ag \setminus A$ to achieve $\bigvee_{i \in A} \neg \AS(\gamma_i)$. This can be done by taking the product of the original game $\Game$ and the finite state transducers representing $\vec{\sigma}_{A}'$, and then model checking the resulting MDP $\MDP_{-A}$, in which $\Ag \setminus A$ is viewed as a single agent, with respect to $\NZ(\neg \gamma_i)$ (line \ref{alg:den_dev1}), which is in \twoexptime and hence so is \textsc{Beneficial Deviation}. 

	To see that this bound is tight note that we can reduce qualitative \LTL model checking on MDPs to \textsc{Beneficial Deviation}. Given an MDP $\MDP$ and \LTL formula $\varphi$, let $\MC^\varphi$ be a minimal (deterministic) MC satisfying $\varphi$ and similarly for $\MC^{\neg\varphi}$. We then form the CSGA $\CGModel$ shown in Figure \ref{fig:ben_dev_reduction}, where player 1 has control over all the actions in $\MDP$, and the resulting game $\Game = (\CGModel, (\ltlnext\varphi, \neg\ltlnext\varphi), \lambda)$ where $\lambda$ is the combination of the labelling functions for $\MDP$, $\MC^\varphi$, and $\MC^{\neg\varphi}$ with $\lambda(s_0) = \varnothing$. Let $ \vec{\sigma} $ be such that $\sigma_1(s_0) = a$ and $\sigma_2(s_0) = b$, and for $A = \{2\}$ let $\vec{\sigma}_{A}'$ be the same as $ \vec{\sigma} $ except for having $\sigma_2(s_0) = \bar{b}$. 

	It can be seen immediately both that $\Game$, $\vec{\sigma}$, and $ \vec{\sigma}_{A}' $ are of size polynomial in the original input $\MDP$ and $\varphi$, and that $ \vec{\sigma}_{A}' $ is a beneficial deviation from $ \vec{\sigma} $ if and only if it is not the case that there exists a strategy in $\MDP$ satisfying $\AS(\varphi)$; if there was then player 1 could switch to such a strategy and play $\bar{a}$ in $s_0$ in order to ensure that the coalition $\{2\}$ does not achieve their goal.
\end{proof}

\begin{algorithm}[t]
	\caption{\textsc{Beneficial Deviation}}
	\begin{algorithmic}[1]
		\Input $\Game$, $\vec{\sigma}$, $ \vec{\sigma}_{A}' $
		\State build $\MDP_{-A}$ from $\Game$ and $ \vec{\sigma}_{A}' $
		\For{$i \in A$}
			\If{$\vec{\sigma} \models \AS(\gamma_i)$} \label{alg:ben_dev0}
				\State \textbf{return} ``no''
			\EndIf
			\If{$\MDP_{-A} \models \NZ(\neg \gamma_i)$} \label{alg:den_dev1}
				\State \textbf{return} ``no''
			\EndIf
		\EndFor
		\Return ``yes''
	\end{algorithmic}
	\label{alg:ben_dev}
\end{algorithm}

%\newpage 
%\section{Special cases}
%\begin{itemize}
%\item memoryless strategies 
%\item one/two-player games 
%\item turn-based games 
%\item Simpler TL goals: e.g., omega-regular given by GF fragment. 
%\item RL agents: e.g., Team games (identical payoff/TL goal). 
%\end{itemize}

%\newpage 
\section{Discussion and Related Work}
\label{secn:conc}

Our results account for cooperative and non-cooperative settings, and required the development of new techniques with respect to previous work. In particular, most verification techniques for deterministic systems, many of which are used in the context of rational verification, no longer apply when probabilistic behaviour is allowed. We then conclude with a more detailed comparison and analysis against previous work on rational verification and probabilistic systems. 

%To the best of our knowledge, this is the first paper that considers the rational verification problem for probabilistic systems in which players' {\em preferences} can be expressed using general \LTL goals, strategies may have access to {\em infinite} memory, interactions can take place {\em concurrently}, and plays may last for an infinite number of rounds, that is, have an {\em infinite horizon}. All previous work, in the probabilistic setting, fails to have at least one of these features arising in full interplay with the others, making our framework the most complex so far developed from a theoretical point of view. 

\subsubsection{Special Cases}
Our results pertain to general-sum multiplayer games with perfect information and memoryful strategies. A few observations can be made as a result. The proofs for hardness of most of our results show that the problems studied here remain in \twoexptime even in the single-player case. The same is true if we consider zero-sum games, which are frequently used in formal verification in the two-player setting. This shows, in turn, that the \twoexptime results hold regardless of the number of players (as long is it is more than one) or how antagonistic they are.

A less obvious situation is when strategies are restricted or when the game has multiple players, but each control their own set of states -- a multiplayer turn-based game. Concrete results in these cases are yet to be obtained (for instance, for memoryless or finite-memory strategies), and therefore have been left as directions for future work. 

Another special case is where players possess simpler temporal logic goals. It is known that two-player games with goals expressed using various kinds of \LTL fragments are \twoexptime-complete even for deterministic systems~\cite{AlurTM03}. However, if only one player is allowed in the game, the problem can have a significantly lower complexity (\np or \exptime) in case of qualitative probabilistic model checking~\cite{KiniV17}. A much different complexity landscape is found when the quantitative probabilistic setting is considered, with similar decision problems having much higher complexity in the quantitative setting, and requiring, yet again, a different toolset of verification techniques. 

\subsubsection{On Rational Verification}
Most relevant related work on rational verification has focused on deterministic systems, leaving all reasoning about stochastic systems largely overlooked. One important question, not discussed in the present paper, is the problem of whether a game has at least one equilibrium: the \textsc{Non-Emptiness} problem in rational verification. This problem always has a positive answer in the cooperative case -- see \cite{GKW2019}, the argument therein also extends to the stochastic setting considered in the current work -- and in the deterministic, non-cooperative case is solvable in \twoexptime when considering arbitrary \LTL goals and strategies. In the stochastic, non-cooperative setting, however, it is an open problem and known results on game theory and games with probabilistic behaviour do not seem to provide an answer. 

Two important types of games in the rational verification literature for multi-agent systems are the case where players control Boolean variables, as in an iterated version of a Boolean game~\cite{ibgs}, and the case where systems are succinctly represented using a guarded command-like language for multi-agent system specifications~\cite{alur:99a}. While all of these problems are also in \twoexptime in the non-stochastic setting, they require considerably different proof techniques for their solutions, typically resorting to the use of logics for strategic reasoning~\cite{MogaveroMPV14}, automata~\cite{FismanKL10}, or reductions to other game representations~\cite{GutierrezNPW20,GutierrezNPW18,BouyerBMU15}. In the probabilistic setting considered here, several new constructions and proof techniques were required, including, {\em e.g.}, the need for infinite-memory strategies to achieve optimal behaviour. 

\subsubsection{Verification of Probabilistic Systems}
Probabilistic systems have been studied extensively, but mostly without a focus on rational behaviour as considered in this work. Instead, the traditional verification approaches to probabilistic systems have considered zero-sum games, $\omega$-regular goals, and many types of winning, of which almost-sure is just one of many possibilities. A comprehensive survey of main results in this area can be found in~\cite{CH2012}. 
Of the many studies on the analysis and verification of probabilistic systems, the work in~\cite{KwiatkowskaNPS20,KwiatkowskaN0S19} is the closest to that presented here. In common with~\cite{KwiatkowskaNPS20,KwiatkowskaN0S19}, we use CSGs as the underlying model and focus on properties satisfied in equilibrium. However, several aspects of our work are different. On one hand, we model players' preferences using \LTL formulae, allow infinite-horizon plays unrestrictedly, provide optimal complexity results for key decision problems, and look at different game-theoretic solution concepts; in fact, no other paper investigates the core for CSGs. On the other hand, we do not consider probabilistic reasoning in the quantitative setting, and do not have a practical implementation. %, which we discuss next.

\subsubsection{Practical Implementations}
So far, only PRISM-games can be used to verify the satisfaction of properties in equilibrium in CSGs, and until recently, only games without concurrency were supported~\cite{KwiatkowskaN0S20}. The current PRISM-games implementation supports concurrency, but verification is restricted to games with a finite horizon, while the procedures we have developed consider plays with an infinite horizon and strategies having access to infinite memory. In addition, PRISM-games considers a non-cooperative solution concept different from Nash equilibrium and does not support cooperative solution concepts.

The closest implementation to the work in this paper, but in a non-probabilistic setting, is EVE~\cite{GutierrezNPW18,GutierrezNPW20}, one of the most efficient software verification tools for the analysis and verification of temporal logic properties of multi-agent systems. An avenue for future work on the practical side is to extend the functionalities of EVE to account for the more complex probabilistic framework we have studied here, both for cooperative and for non-cooperative games. MCMAS~\cite{LomuscioQR17}, a verification tool for multi-agent systems, also provides some support to model check logics for strategic reasoning, some of which can express both Nash equilibrium and the core; however, at the time of writing, an implementation that can account for the kind of probabilistic systems we have considered here is not available.

\subsubsection{Future Work} 
As pointed out before, a few problems seem to lead to interesting avenues for future research. On the practical side, just discussed, the immediate work to do would be to implement the algorithms herein proposed -- an powerful way to extend the current capabilities of, say, EVE's verification engine which at present does not support any kind of probabilistic reasoning. On the other hand, on the theory side, we would like to understand better two specific problems: firstly, \textsc{Non-Emptiness} in the stochastic, non-cooperative setting, and secondly, whether our results can be extended to the probabilistic quantitative setting.

\section*{Acknowledgments}

Lewis Hammond acknowledges the support of an EPSRC Doctoral Training Partnership studentship (Reference: 2218880). Anthony Lin and Muhammad Najib acknowledge the support of ERC Starting Grant 759969 (AV-SMP) and Max-Planck Fellowship. Michael Wooldridge was supported by JP Morgan and the Alan Turing Institute. 
%  the \grantsponsor{GS501100001809}{JP Morgan and The Alan Turing Institute}{http://dx.doi.org/10.13039/501100001809} under Grant
%  No.:~\grantnum{GS501100001809}{61273304}
%  and~\grantnum[http://www.nnsf.cn/youngscientsts]{GS501100001809}{Young
%    Scientists' Support Program}.

\appendix

%% The file kr.bst is a bibliography style file for BibTeX 0.99c
\bibliographystyle{kr}
\bibliography{biblio}

\begin{thebibliography}{}

\bibitem[\protect\citeauthoryear{Abdulla \bgroup et al\mbox.\egroup
  }{2008}]{SLCS-abdulla}
Abdulla, P.~A.; Henda, N.~B.; de~Alfaro, L.; Mayr, R.; and Sandberg, S.
\newblock 2008.
\newblock Stochastic games with lossy channels.
\newblock In Amadio, R.~M., ed., {\em FoSSaCS}, volume 4962 of {\em Lecture
  Notes in Computer Science},  35--49.
\newblock Springer.

\bibitem[\protect\citeauthoryear{Alur and Henzinger}{1999}]{alur:99a}
Alur, R., and Henzinger, T.
\newblock 1999.
\newblock Reactive modules.
\newblock {\em Formal Methods in System Design} 15(1):7--48.

\bibitem[\protect\citeauthoryear{Alur, Torre, and Madhusudan}{2003}]{AlurTM03}
Alur, R.; Torre, S.~L.; and Madhusudan, P.
\newblock 2003.
\newblock Playing games with boxes and diamonds.
\newblock In Amadio, R.~M., and Lugiez, D., eds., {\em {CONCUR}}, volume 2761
  of {\em LNCS},  127--141.
\newblock Springer.

\bibitem[\protect\citeauthoryear{Baier and Katoen}{2008}]{BK08}
Baier, C., and Katoen, J.-P.
\newblock 2008.
\newblock {\em Principles of Model Checking (Representation and Mind Series)}.
\newblock The MIT Press.

\bibitem[\protect\citeauthoryear{Baier, Bertrand, and
  Schnoebelen}{2007}]{SLCS-bertrand}
Baier, C.; Bertrand, N.; and Schnoebelen, P.
\newblock 2007.
\newblock Verifying nondeterministic probabilistic channel systems against
  {\(\omega\)}-regular linear-time properties.
\newblock {\em {ACM} Trans. Comput. Log.} 9(1):5.

\bibitem[\protect\citeauthoryear{Berthon, Guha, and Raskin}{2020}]{BGR2020}
Berthon, R.; Guha, S.; and Raskin, J.-F.
\newblock 2020.
\newblock Mixing probabilistic and non-probabilistic objectives in markov
  decision processes.
\newblock In {\em Proceedings of the 35th Annual ACM/IEEE Symposium on Logic in
  Computer Science}, LICS '20,  195–208.
\newblock New York, NY, USA: Association for Computing Machinery.

\bibitem[\protect\citeauthoryear{Bouyer \bgroup et al\mbox.\egroup
  }{2015}]{BouyerBMU15}
Bouyer, P.; Brenguier, R.; Markey, N.; and Ummels, M.
\newblock 2015.
\newblock Pure nash equilibria in concurrent deterministic games.
\newblock {\em Log. Methods Comput. Sci.} 11(2).

\bibitem[\protect\citeauthoryear{Chatterjee and Henzinger}{2012}]{CH2012}
Chatterjee, K., and Henzinger, T.~A.
\newblock 2012.
\newblock A survey of stochastic $\omega$-regular games.
\newblock {\em Journal of Computer and System Sciences} 78(2):394 -- 413.
\newblock Games in Verification.

\bibitem[\protect\citeauthoryear{Courcoubetis and Yannakakis}{1995}]{CY95}
Courcoubetis, C., and Yannakakis, M.
\newblock 1995.
\newblock The complexity of probabilistic verification.
\newblock {\em J. ACM} 42(4):857–907.

\bibitem[\protect\citeauthoryear{{de Alfaro} and
  Henzinger}{2000}]{deAlfaro2000}
{de Alfaro}, L., and Henzinger, T.~A.
\newblock 2000.
\newblock Concurrent omega-regular games.
\newblock In {\em Proceedings of the 15th Annual IEEE Symposium on Logic in
  Computer Science}, LICS '00,  141.
\newblock USA: IEEE Computer Society.

\bibitem[\protect\citeauthoryear{Fisman, Kupferman, and
  Lustig}{2010}]{FismanKL10}
Fisman, D.; Kupferman, O.; and Lustig, Y.
\newblock 2010.
\newblock Rational synthesis.
\newblock In Esparza, J., and Majumdar, R., eds., {\em {TACAS}}, volume 6015 of
  {\em LNCS},  190--204.
\newblock Springer.

\bibitem[\protect\citeauthoryear{Gutierrez \bgroup et al\mbox.\egroup
  }{2018}]{GutierrezNPW18}
Gutierrez, J.; Najib, M.; Perelli, G.; and Wooldridge, M.~J.
\newblock 2018.
\newblock {EVE:} {A} tool for temporal equilibrium analysis.
\newblock In Lahiri, S.~K., and Wang, C., eds., {\em {ATVA}}, volume 11138 of
  {\em LNCS},  551--557.
\newblock Springer.

\bibitem[\protect\citeauthoryear{Gutierrez \bgroup et al\mbox.\egroup
  }{2020}]{GutierrezNPW20}
Gutierrez, J.; Najib, M.; Perelli, G.; and Wooldridge, M.~J.
\newblock 2020.
\newblock Automated temporal equilibrium analysis: Verification and synthesis
  of multi-player games.
\newblock {\em Artif. Intell.} 287:1--70.

\bibitem[\protect\citeauthoryear{Gutierrez, Harrenstein, and
  Wooldridge}{2015a}]{GHW15}
Gutierrez, J.; Harrenstein, P.; and Wooldridge, M.
\newblock 2015a.
\newblock {Expresiveness and Complexity Results for Strategic Reasoning}.
\newblock In {\em 26th International Conference on Concurrency Theory (CONCUR
  2015)}.

\bibitem[\protect\citeauthoryear{Gutierrez, Harrenstein, and
  Wooldridge}{2015b}]{ibgs}
Gutierrez, J.; Harrenstein, P.; and Wooldridge, M.
\newblock 2015b.
\newblock Iterated boolean games.
\newblock {\em Inf. Comput.} 242(C):53–79.

\bibitem[\protect\citeauthoryear{Gutierrez, Harrenstein, and
  Wooldridge}{2017}]{GutierrezHW17}
Gutierrez, J.; Harrenstein, P.; and Wooldridge, M.~J.
\newblock 2017.
\newblock From model checking to equilibrium checking: Reactive modules for
  rational verification.
\newblock {\em Artif. Intell.} 248:123--157.

\bibitem[\protect\citeauthoryear{Gutierrez, Kraus, and
  Wooldridge}{2019}]{GKW2019}
Gutierrez, J.; Kraus, S.; and Wooldridge, M.
\newblock 2019.
\newblock Cooperative concurrent games.
\newblock AAMAS '19,  1198–1206.
\newblock Richland, SC: International Foundation for Autonomous Agents and
  Multiagent Systems.

\bibitem[\protect\citeauthoryear{Kini and Viswanathan}{2017}]{KiniV17}
Kini, D., and Viswanathan, M.
\newblock 2017.
\newblock Complexity of model checking mdps against {LTL} specifications.
\newblock In Lokam, S.~V., and Ramanujam, R., eds., {\em {FSTTCS}}, volume~93
  of {\em LIPIcs},  35:1--35:13.
\newblock Schloss Dagstuhl - Leibniz-Zentrum f{\"{u}}r Informatik.

\bibitem[\protect\citeauthoryear{Kwiatkowska \bgroup et al\mbox.\egroup
  }{2019}]{KwiatkowskaN0S19}
Kwiatkowska, M.; Norman, G.; Parker, D.; and Santos, G.
\newblock 2019.
\newblock Equilibria-based probabilistic model checking for concurrent
  stochastic games.
\newblock In ter Beek, M.~H.; McIver, A.; and Oliveira, J.~N., eds., {\em
  {FM}}, volume 11800 of {\em LNCS},  298--315.
\newblock Springer.

\bibitem[\protect\citeauthoryear{Kwiatkowska \bgroup et al\mbox.\egroup
  }{2020a}]{KwiatkowskaNPS20}
Kwiatkowska, M.; Norman, G.; Parker, D.; and Santos, G.
\newblock 2020a.
\newblock Multi-player equilibria verification for concurrent stochastic games.
\newblock In Gribaudo, M.; Jansen, D.~N.; and Remke, A., eds., {\em {QEST}},
  volume 12289 of {\em LNCS},  74--95.
\newblock Springer.

\bibitem[\protect\citeauthoryear{Kwiatkowska \bgroup et al\mbox.\egroup
  }{2020b}]{KwiatkowskaN0S20}
Kwiatkowska, M.; Norman, G.; Parker, D.; and Santos, G.
\newblock 2020b.
\newblock Prism-games 3.0: Stochastic game verification with concurrency,
  equilibria and time.
\newblock In Lahiri, S.~K., and Wang, C., eds., {\em {CAV}}, volume 12225 of
  {\em LNCS},  475--487.
\newblock Springer.

\bibitem[\protect\citeauthoryear{Lehmann and Rabin}{1981}]{LR81}
Lehmann, D., and Rabin, M.
\newblock 1981.
\newblock On the advantage of free choice: {A} symmetric and fully distributed
  solution to the dining philosophers problem (extended abstract).
\newblock In {\em POPL},  133--138.

\bibitem[\protect\citeauthoryear{Leng{\'{a}}l \bgroup et al\mbox.\egroup
  }{2017}]{fairy-tale}
Leng{\'{a}}l, O.; Lin, A.~W.; Majumdar, R.; and R{\"{u}}mmer, P.
\newblock 2017.
\newblock Fair termination for parameterized probabilistic concurrent systems.
\newblock In {\em {TACAS}},  499--517.

\bibitem[\protect\citeauthoryear{Lin and R{\"{u}}mmer}{2016}]{LR16}
Lin, A.~W., and R{\"{u}}mmer, P.
\newblock 2016.
\newblock Liveness of randomised parameterised systems under arbitrary
  schedulers.
\newblock In {\em {CAV}},  112--133.

\bibitem[\protect\citeauthoryear{Lomuscio, Qu, and
  Raimondi}{2017}]{LomuscioQR17}
Lomuscio, A.; Qu, H.; and Raimondi, F.
\newblock 2017.
\newblock {MCMAS:} an open-source model checker for the verification of
  multi-agent systems.
\newblock {\em Int. J. Softw. Tools Technol. Transf.} 19(1):9--30.

\bibitem[\protect\citeauthoryear{Mogavero \bgroup et al\mbox.\egroup
  }{2014}]{MogaveroMPV14}
Mogavero, F.; Murano, A.; Perelli, G.; and Vardi, M.~Y.
\newblock 2014.
\newblock Reasoning about strategies: On the model-checking problem.
\newblock {\em {ACM} Trans. Comput. Log.} 15(4):34:1--34:47.

\bibitem[\protect\citeauthoryear{Norman}{2004}]{Norman04}
Norman, G.
\newblock 2004.
\newblock Analysing randomized distributed algorithms.
\newblock In {\em Validation of Stochastic Systems - {A} Guide to Current
  Research},  384--418.

\bibitem[\protect\citeauthoryear{Osborne and Rubinstein}{1994}]{OR94}
Osborne, M.~J., and Rubinstein, A.
\newblock 1994.
\newblock {\em A Course in Game Theory}.
\newblock MIT Press.

\bibitem[\protect\citeauthoryear{{Piterman}}{2006}]{Piterman2007}
{Piterman}, N.
\newblock 2006.
\newblock From nondeterministic buchi and streett automata to deterministic
  parity automata.
\newblock In {\em 21st Annual IEEE Symposium on Logic in Computer Science
  (LICS'06)},  255--264.

\bibitem[\protect\citeauthoryear{Pnueli}{1977}]{pnueli:77a}
Pnueli, A.
\newblock 1977.
\newblock The temporal logic of programs.
\newblock In {\em FOCS},  46--57.
\newblock IEEE.

\bibitem[\protect\citeauthoryear{{Vardi}}{1985}]{vardi85}
{Vardi}, M.~Y.
\newblock 1985.
\newblock Automatic verification of probabilistic concurrent finite state
  programs.
\newblock In {\em 26th Annual Symposium on Foundations of Computer Science
  (sfcs 1985)},  327--338.

\bibitem[\protect\citeauthoryear{Wooldridge \bgroup et al\mbox.\egroup
  }{2016}]{WooldridgeGHMPT16}
Wooldridge, M.~J.; Gutierrez, J.; Harrenstein, P.; Marchioni, E.; Perelli, G.;
  and Toumi, A.
\newblock 2016.
\newblock Rational verification: From model checking to equilibrium checking.
\newblock In Schuurmans, D., and Wellman, M.~P., eds., {\em {AAAI}},
  4184--4191.
\newblock {AAAI} Press.

\end{thebibliography}

\end{document}